\documentclass[a4paper,twoside]{article}
\usepackage[INRIA]{lipinria}

\usepackage[latin1]{inputenc}
\usepackage[T1]{fontenc}
\usepackage{amsfonts,amsmath,amssymb,amsthm,amstext,latexsym}
\usepackage{figlatex}
\graphicspath{{Figures/}}
\usepackage{xspace}
\usepackage{subfigure}
\usepackage[ruled,vlined]{algorithm2e}
\usepackage{url}
\usepackage{relsize}
\usepackage{enumerate}
\usepackage{colortbl}
\usepackage{stmaryrd}
\usepackage{boxedminipage}

\newtheorem{proposition}{Proposition}

\AtBeginDocument{}

\newcommand{\A}{\ensuremath{\mathcal{A}}\xspace}
\newcommand{\B}{\ensuremath{\mathcal{B}}\xspace}
\newcommand{\C}{\ensuremath{\mathcal{C}}\xspace}

\newcommand{\MC}[1]{{\boldmath$\mathcal{C}_{#1}$}}
\newcommand{\rrr}{r}
\newcommand{\sss}{s}
\newcommand{\ttt}{t}

\newcommand{\calc}{w} 
\newcommand{\comm}{c} 
\newcommand{\mem}{m} 
\newcommand{\nbp}{p} 
\newcommand{\CCRB}{\text{CCR}} 
\newcommand{\usedp}{\ensuremath{\mathfrak{P}}} 
\newcommand{\usedq}{\ensuremath{\mathfrak{Q}}} 
\newcommand{\block}{\ensuremath{\mathbf{C}}} 

\newcommand{\mm}{\ensuremath{\mu}} 

\newcommand{\aold}{\ensuremath{\alpha_{old}}}
\newcommand{\betaold}{\ensuremath{\beta_{old}}}
\newcommand{\cold}{\ensuremath{\gamma_{old}}}
\newcommand{\anew}{\ensuremath{\alpha_{recv}}}
\newcommand{\bnew}{\ensuremath{\beta_{recv}}}
\newcommand{\cnew}{\ensuremath{\gamma_{recv}}}
\newcommand{\csend}{\ensuremath{\gamma_{send}}}

\newcommand{\la}{\leftarrow}

\newcommand{\ratio}{\textsf{ratio}}
\newcommand{\nextt}{\textsf{next}}
\newcommand{\totwork}{\textsf{total-work}}
\newcommand{\complet}{\textsf{completion-time}}
\newcommand{\ready}{\textsf{ready}}
\newcommand{\nblock}{\textsf{nb-block}}
\newcommand{\nbcol}{\textsf{nb-column}}

\begin{document}

\RRItheme{\THNum}

\RRIprojet{GRAAL}

\RRInumber{2006-39}

\RRItitle{Revisiting Matrix Product on Master-Worker Platforms}
\RRIthead{Revisiting Matrix Product on Master-Worker Platforms}

\RRItitre{Produit de matrice sur plate-forme maître-esclave}

\RRIauthor{Jack Dongarra\and Jean-François Pineau\and Yves Robert\and Zhiao Shi\and Frédéric Vivien}
\RRIahead{J. Dongarra\and J.-F. Pineau\and Y. Robert\and Z. Shi\and F. Vivien}

\RRIdate{December 2006}

\RRIkeywords{Matrix product, LU decomposition, Master-worker platform, Heterogeneous platforms, Scheduling}

\RRImotscles{Produit de matrices, Décomposition LU, Plates-formes maître-esclaves, Plates-formes hétérogènes, Ordonnancement}

\RRIabstract{
This paper is aimed at designing efficient
parallel matrix-product algorithms for  heterogeneous master-worker platforms.
While matrix-product is well-understood for \emph{homogeneous
2D-arrays of processors} (e.g., Cannon algorithm and ScaLAPACK outer
product algorithm), there are three
key hypotheses that render our work original and innovative:\\
- \emph{Centralized data.} We assume that all matrix files originate from,
and must be returned to, the master. The master distributes both data
and computations to the workers (while in ScaLAPACK, input and output matrices are
initially distributed among participating resources). Typically, our approach
is useful in the context of speeding up MATLAB or SCILAB clients running on a server
(which acts as the master and initial repository of files).\\
- \emph{Heterogeneous star-shaped platforms.} We target fully heterogeneous platforms,
 where computational resources have different computing powers. Also, the workers are connected to
 the master by links of different capacities. This framework is realistic when deploying
 the application from the server, which is responsible for enrolling authorized resources.\\
- \emph{Limited memory.} Because we investigate the parallelization of large problems,
 we cannot assume that full matrix panels can be stored in the worker memories and re-used for
 subsequent updates (as in ScaLAPACK). The amount of memory available in each worker is
 expressed as a given number $\mem_i$ of buffers, where a buffer can store a square block
 of matrix elements. The size  $q$ of these square blocks  is chosen so as to harness
 the power of Level 3 BLAS routines: $q=80$ or $100$ on most platforms.\\

\medskip
We have devised efficient algorithms for resource selection (deciding which workers to enroll)
and communication ordering (both for input and result messages),
and we report a set of numerical experiments on various
platforms at École Normale Supérieure de Lyon and the University of Tennessee.
However, we point out that in this first version of the report, experiments are limited to
homogeneous platforms.}

\RRIresume{Ce papier a pour objectif la définition d'algorithmes
  efficaces pour le produit de matrices en parallèle sur plate-formes
  maître-esclaves hétérogènes. Bien que le produit de matrices soit
  bien compris pour des \emph{grilles bi-dimensionnelles de
    processeurs homogènes} (cf. l'algorithme de Cannon et le produit
  externe de ScaLAPACK), trois hypothèses rendent notre travail
  original:\\
  - \emph{Données centralisées.} Nous supposons que toutes les
  matrices résident originellement sur le maître, et doivent y être
  renvoyées. Le maître distribue données et calculs aux esclaves
  (alors que dans ScaLAPACK, les matrices initiales et résultats sont
  initiallement distribuées aux processeurs participant). Typiquement,
  notre approche est justifiée dans le contexte de l'accélération de
  clients MATLAB ou SCILAB s'exécutant sur un serveur (qui se comporte
  comme le maître et détient initiallement les données).\\
  
  - \emph{Plates-formes hétérogènes en étoile.} Nous nous intéressons
  à des plates-formes complètement hétérogènes dont les ressources de
  calculs ont des puissances de calcul différentes et dont les esclaves
  sont reliés au maître par des liens de capacités différentes. Ce
  cadre de travail est réaliste quand l'application est déployée à
  partir du serveur qui est responsable de l'enrôlement des ressources
  nécessaires.\\

  - \emph{Mémoire bornée.} Comme nous nous intéressons à la
  parallélisation de gros problèmes, nous ne pouvons pas supposer que
  toutes les sous-matrices peuvent être stockées dans la mémoire de
  chaque esclave pour être éventuellement réutilisée ultérieurement
  (comme c'est le cas dans ScaLAPACK). La quantité de mémoire
  disponible sur un esclave donné est exprimé comme un nombre $\mem_i$
  de tampons, où un tampon peut exactement contenir un bloc carré
  d'éléments de matrice. La taille $q$ de ces blocs carrés est choisie
  afin de pouvoir tirer parti de la puissance des routines BLAS de
  niveau 3: $q=80$ ou $100$ sur la plupart des plates-formes.\\

  \medskip Nous avons défini des algorithmes efficaces pour la
  sélection de ressources (pour décider quel(s) esclave(s) utiliser)
  et l'ordonnancement des communications (envoi de données et
  récupérations de résultats), et nous rapportons un ensemble
  d'expériences sur des plates-formes à l'École normale supérieure de
  Lyon et à l'Université du Tennessee. Nous faisons cependant
  remarquer que dans la première version de ce rapport les expériences
  ne concernent que des plates-formes homogènes.}

\RRImaketitle

\section{Introduction}
\label{sec.intro}

Matrix product is a key computational kernel in many scientific
applications, and it has been extensively studied on parallel
architectures. Two well-known parallel versions are Cannon's
algorithm~\cite{Cannon69} and the ScaLAPACK outer product
algorithm~\cite{Scalapack97}.  Typically, parallel implementations
work well on 2D processor grids, because the input matrices are sliced
horizontally and vertically into square blocks that are mapped
one-to-one onto the physical resources; several communications can
take place in parallel, both horizontally and vertically. Even better,
most of these communications can be overlapped with (independent)
computations.  All these characteristics render the matrix product
kernel quite amenable to an efficient parallel implementation on 2D
processor grids.

However, current architectures typically take the form of
heterogeneous clusters, which are composed of heterogeneous computing
resources, interconnected by a \emph{sparse} network: there are no
direct links between any pair of processors. Instead, messages from
one processor to another are routed via several links, likely to have
different capacities. Worse, congestion will occur when two messages,
involving two different sender/receiver pairs, collide because a
same physical link happens to belong to the two routing paths.
Therefore, an accurate estimation of the communication cost requires a
precise knowledge of the underlying target platform. In addition, it
becomes necessary to include the cost of both the initial distribution of
the matrices to the processors and of collecting back the results.
These input/output operations have always been neglected in the
analysis of the conventional algorithms. This is because only $O(n^2)$
coefficients need to be distributed in the beginning, and gathered at
the end, as opposed to the $O(n^3)$ computations to be performed
(where $n$ is the problem size). The assumption that these
communications can be ignored could have made sense on dedicated
processor grids like, say, the Intel Paragon, but it is no longer
reasonable on heterogeneous platforms.

There are two possible approaches to tackle the parallelization of
matrix product on heterogeneous clusters when aiming at reusing the 2D
processor grid strategy. The first (drastic) approach is to
\emph{ignore} communications. The objective is then to load-balance
computations as evenly as possible on a heterogeneous 2D processor
grid. This corresponds to arranging the $n$ available resources as a
(virtual) 2D grid of size $p \times q$ (where $p.q \leq n$) so that
each processor receives a share of the work, i.e., a rectangle, whose
area is proportional to its relative computing speed.  There are many
processor arrangements to consider, and determining the optimal one is
a highly combinatorial problem, which has been proven NP-complete
in~\cite{ieeeTC2001}. In fact, because of the geometric constraints
imposed by the 2D processor grid, a perfect load-balancing can only be
achieved in some very particular cases.

The second approach is to relax the geometric constraints imposed by a
2D processor grid. The idea is then to search for a 2D partitioning of
the input matrices into rectangles that will be mapped one-to-one
onto the processors. Because the 2D partitioning now is irregular (it
is no longer constrained to a 2D grid), some processors may well have
more than four neighbors. The advantage of this approach is that a
perfect load-balancing is always possible; for instance partitioning
the matrices into horizontal slices whose vertical dimension is
proportional to the computing speed of the processors always leads to
a perfectly balanced distribution of the computations. The objective
is then to minimize the total cost of the communications. However, it
is very hard to accurately predict this cost. Indeed, the processor
arrangement is virtual, not physical: as explained above, the
underlying interconnection network is not expected to be a complete
graph, and communications between neighbor processors in the
arrangement are likely to be realized via several physical links
constituting the communication path. The actual repartition of the
physical links across all paths is hard to predict, but contention is
almost certain to occur. This is why a natural, although pessimistic
assumption, to estimate the communication cost, is to assume that all
communications in the execution of the algorithm will be implemented
sequentially. With this hypothesis, minimizing the total communication
cost amounts to minimizing the total communication volume.
Unfortunately, this problem has been shown NP-complete as
well~\cite{ieeeTPDS2001}. Note that even under the optimistic
assumption that all communications at a given step of the algorithm
can take place in parallel, the problem remains
NP-complete~\cite{algorithmica}.

In this paper, we do not try to adapt the 2D processor grid strategy
to heterogeneous clusters. Instead, we adopt a realistic application
scenario, where input files are read from a fixed repository (disk on
a data server).  Computations will be delegated to available resources
in the target architecture, and results will be returned to the
repository.  This calls for a master-worker paradigm, or more
precisely for a computational scheme where the master (the processor
holding the input data) assigns computations to other resources, the
workers.  In this centralized approach, all matrix files originate
from, and must be returned to, the master. The master distributes both
data and computations to the workers (while in ScaLAPACK, input and
output matrices are supposed to be equally distributed among
participating resources beforehand).  Typically, our approach is
useful in the context of speeding up MATLAB or SCILAB clients running
on a server (which acts as the master and initial repository of
files).

We target fully heterogeneous master-worker platforms, where
computational resources have different computing powers. Also, the
workers are connected to the master by links of different capacities.
This framework is realistic when deploying the application from the
server, which is responsible for enrolling authorized resources.

Finally, because we investigate the parallelization of large problems,
we cannot assume that full matrix panels can be stored in worker
memories and re-used for subsequent updates (as in ScaLAPACK).  The
amount of memory available in each worker is expressed as a given
number $\mem_i$ of buffers, where a buffer can store a square block of
matrix elements. The size $q$ of these square blocks is chosen so as
to harness the power of Level 3 BLAS routines: $q=80$ or $100$ on most
platforms.

To summarize, the target platform is composed of several workers with
different computing powers, different bandwidth links to/from the
master, and different, limited, memory capacities.  The first problem
is \emph{resource selection}.  Which workers should be enrolled in the
execution?  All of them, or maybe only the faster computing ones, or
else only the faster-communicating ones? Once participating resources
have been selected, there remain several scheduling decisions to take:
how to minimize the number of communications? in which order workers
should receive input data and return results?  what amount of
communications can be overlapped with (independent) computations?  The
goal of this paper is to design efficient algorithms for resource
selection and communication ordering. In addition, we report numerical
experiments on various heterogeneous platforms at the École Normale
Supérieure de Lyon and at the University of Tennessee.

The rest of the paper is organized as follows. In
Section~\ref{sec.framework}, we state the scheduling problem
precisely, and we introduce some notations.  In
Section~\ref{sec.stripes}, we start with a theoretical study of the
simplest version of the problem, without memory limitation, which is
intended to show the intrinsic difficulty of the scheduling problem.
Next, in Section~\ref{sec.volume}, we proceed with the analysis of the
total communication volume that is needed in the presence of memory
constraints, and we improve a well-known bound by
Toledo~\cite{toledo99survey,toledoJPDC}.  We deal with homogeneous
platforms in Section~\ref{sec.homo}, and we propose a scheduling
algorithm that includes resource selection. Section~\ref{sec.hetero}
is the counterpart for heterogeneous platforms, but the algorithms are
much more complicated. In Section~\ref{sec.LU}, we briefly discuss how to extend previous
approaches to LU factorization. We report several MPI experiments in
Section~\ref{sec.MPI}. Section~\ref{sec.related} is devoted to an
overview of related work. Finally, we state some concluding remarks in
Section~\ref{sec.conclusion}.

\section{Framework}
\label{sec.framework}

In this section we formally state our hypotheses on the application
(Section~\ref{sec.appli}) and on the target platform
(Section~\ref{sec.platform}).
\begin{figure}[htb]
\begin{minipage}[b]{0.48\linewidth}
  \centering
  \includegraphics[width=0.9\textwidth]{matmult.fig}
  \caption{Partition of the three matrices $\A$, $\B$, and $\C$.}
  \label{fig.partition}
\end{minipage}
\hfill
\begin{minipage}[b]{0.48\linewidth}
      \centering
      \includegraphics[width=0.9\textwidth]{platform.fig}
      \caption{A fully heterogeneous master-worker platform.}
      \label{fig.platform}
\end{minipage}
\end{figure}



\subsection{Application}
\label{sec.appli}

We deal with the computational kernel $\C \leftarrow \C + \A \times
\B$. We partition the three matrices $\A$, $\B$, and $\C$ as
illustrated in Figure~\ref{fig.partition}. More precisely:

\begin{itemize}
\item We use a block-oriented approach. The atomic elements that we
  manipulate are not matrix coefficients but instead square
  \emph{blocks} of size $q \times q$ (hence with $q^2$ coefficients).
  This is to harness the power of Level 3 BLAS
  routines~\cite{BlackfordCC96}. Typically, $q=80$ or $100$ when using
  ATLAS-generated routines~\cite{atlas_sc98}.

\item The input matrix $\A$ is of size $n_{\A} \times n_{\A\B}$:\\
  - we split $\A$ into $\rrr$ horizontal stripes $\A_i$, $1 \leq i \leq \rrr$, where $\rrr = n_{\A} / q$;\\
  - we split each stripe $\A_i$ into $\ttt$ square $q \times q$ blocks
  $\A_{i,k}$, $1 \leq k \leq \ttt$, where $\ttt = n_{\A\B} / q$.

\item The input matrix $\B$ is of size $n_{\A\B} \times n_{\B}$:\\
  - we split $\B$ into $\sss$ vertical stripes $\B_j$, $1 \leq j \leq \sss$, where $\sss = n_{\B}/ q$;\\
  - we split stripe $\B_j$ into $\ttt$ square $q \times q$ blocks
  $\B_{k,j}$, $1 \leq k \leq \ttt$.

\item We compute $\C = \C + \A \times \B$. Matrix $\C$ is accessed
  (both for input and output) by square $q \times q$ blocks
  $\C_{i,j}$, $1 \leq i \leq \rrr$, $1 \leq j \leq \sss$.  There are
  $\rrr \times \sss$ such blocks.

\end{itemize}

We point out that with such a decomposition all stripes and blocks
have same size. This will greatly simplify the analysis of
communication costs.

\subsection{Platform}
\label{sec.platform}

We target a \emph{star network}
$\mathcal{S} = \{ P_0, P_1, P_2, \ldots, P_\nbp \}$, composed of a
master $P_0$ and of $\nbp$ workers $P_i$, $1 \leq i \leq \nbp$ (see Figure~\ref{fig.platform}).
Because we manipulate large data blocks, we adopt a linear cost model,
both for computations and communications (i.e., we neglect start-up overheads).
We have the following notations:
\begin{itemize}
  \item It takes $X. \calc_i$ time-units to execute a task of size $X$ on $P_i$;

  \item It takes $X. \comm_i$ time units for the master $P_0$ to
send a message of size $X$ to $P_i$ or to receive a message of size $X$ from $P_i$.
\end{itemize}

Our star platforms are thus fully heterogeneous, both in terms of
computations and of communications. A fully homogeneous star platform
would be a star platform with identical workers and identical
communication links: $\calc_i = \calc$ and $\comm_i = \comm$ for each
worker $P_i$, $1 \leq i \leq \nbp$.
Without loss of generality, we assume that the master has no
processing
capability (otherwise, add a fictitious extra worker paying no
communication cost to simulate computation at the master).

Next, we need to define the communication model.
We adopt the \emph{one-port} model~\cite{Bhat99efficient,BhatRagra03},
which is defined as follows:
\begin{itemize}
\item  the master can only send data to, and receive data from,
a single worker at a given time-step,
\item  a given worker cannot start execution
before it has terminated the reception of the message from the master;
similarly, it cannot
start sending the results back to the master before finishing the computation.
\end{itemize}
In fact, this \emph{one-port} model naturally comes in two flavors with return messages,
depending upon whether we allow the master to simultaneously send and receive messages
or not. If we do allow for simultaneous sends and receives, we have the \emph{two-port}
model. Here we concentrate on
the true \emph{one-port} model, where the master cannot be enrolled in more than one
communication at any time-step.

The \emph{one-port} model is \emph{realistic}. Bhat, Raghavendra, and
Prasanna~\cite{Bhat99efficient,BhatRagra03} advocate its use because
``current hardware and software do not easily enable multiple messages
to be transmitted simultaneously.'' Even if non-blocking
multi-threaded communication libraries allow for initiating multiple
send and receive operations, they claim that all these operations
``are eventually serialized by the single hardware port to the
network.'' Experimental evidence of this fact has recently been
reported by Saif and Parashar~\cite{SaifPa2004}, who report that
asynchronous MPI sends get serialized as soon as message sizes exceed
a hundred kilobytes. Their result hold for two popular MPI
implementations, MPICH on Linux clusters and IBM MPI on the SP2.  Note
that all the MPI experiments in Section~\ref{sec.MPI} obey the
one-port model.

The one-port model fully accounts for the heterogeneity of the
platform, as each link has a different bandwidth. It generalizes a
simpler model studied by Banikazemi, Moorthy, and
Panda~\cite{Banikazemi98}, Liu~\cite{Liu02}, and Khuller and
Kim~\cite{khuller04soda}. In this simpler model, the communication
time only depends on the sender, not on the receiver. In other words,
the communication speed from a processor to all its neighbors is the
same. This would restrict the study to bus platforms instead of
general star platforms.

Our final assumption is related to memory capacity; we assume that
a worker $P_i$ can only store $\mem_i$ blocks (either from $\A$, $\B$, or $\C$).
For large problems, this memory limitation will considerably impact the design of the
algorithms, as data re-use will be greatly dependent on the amount of available buffers.

\section{Combinatorial complexity of a simple version of the problem}
\label{sec.stripes}

This section is almost a digression; it is devoted to the study of the
simplest variant of the problem.  It is intended to show the intrinsic
combinatorial difficulty of the problem. We make the following
simplifications:
\begin{itemize}
  \item We target a fully homogeneous platform (identical workers and communication links).
  \item We consider only rank-one block updates; in other words, and
    with previous notations, we focus on the case where $\ttt=1$.
  \item Results need not be returned to the master.
  \item Workers have \emph{no} memory limitation; they receive each
    stripe only once and can re-use them for other computations.
\end{itemize}

There are five parameters in the problem; three platform parameters
($\comm$, $\calc$, and the number of workers $\nbp$) and two
application parameters ($\rrr$ and $\sss$).
The scheduling problem amounts to deciding which files should be sent to
which workers and in which order.  A given file may well be sent
several times, to further distribute computations. For instance, a
simple strategy is to partition $\A$ and to duplicate $\B$, i.e., send
each block $\A_i$ only once and each block $\B_j$ $\nbp$ times; all
workers would then be able to work fully in parallel.

\begin{figure}[htb]
      \centering
      \includegraphics[width=0.3\textwidth]{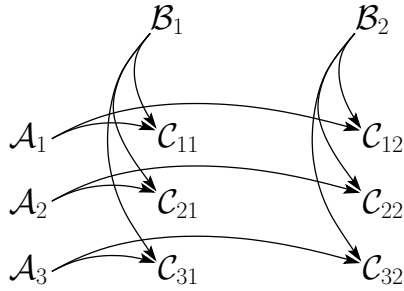}
      \caption{Dependence graph of the problem (with $\rrr=3$ and $\sss=2$).}
      \label{fig.biparti}
\end{figure}

The dependence graph of the problem is depicted in
Figure~\ref{fig.biparti}. It suggests a natural strategy for enabling
workers to start computing as soon as possible. Indeed, the master
should alternate sending $\A$-blocks and $\B$-blocks. Of course it must be decided how many workers to enroll and in which order to
send the blocks to the enrolled workers. But with a single worker, we
can show that the \emph{alternating greedy} algorithm is optimal:

\begin{proposition}
With a single worker, the \emph{alternating greedy} algorithm is optimal.
\end{proposition}

\begin{proof}
In this algorithm, the master sends blocks as soon as possible,
alternating a block of type $\A$ and a block of type $\B$ (and proceeds with the remaining blocks
when one type is exhausted). This strategy maximizes at each step
the total number of tasks that can be processed by the worker. To see this, after $x$ communication steps,
with $y$ files of type $\A$ sent, and $z$ files of type $\B$ sent, where $y+z=x$,
the worker can process at most $y \times z$ tasks. The greedy algorithm enforces $y = \lceil \frac{x}{2} \rceil$
and $z = \lfloor \frac{x}{2} \rfloor$ (as long as $\max(x,y) \leq \min(\rrr,\sss)$, and then
sends the remaining files), hence its optimality.
\end{proof}

Unfortunately, for more than one worker, we did not succeed in determining an optimal algorithm.
There are (at least) two greedy algorithms that can be devised for $\nbp$ workers:

\begin{description}
\item[Thrifty:] This algorithm ``spares'' resources as it aims at keeping each enrolled worker fully active.
It works as follows:
\begin{itemize}
\item Send enough blocks to the first worker so that it is never idle,
\item Send blocks to a second worker during spare communication slots, and
\item Enroll a new worker (and send blocks to it) only if this does not delay previously enrolled workers.
\end{itemize}

\item[Min-min:] This algorithm is based on the well-known min-min heuristic~\cite{MaheswaranAS99}.
At each step, all tasks are considered. For each of them, we compute their possible starting date
on each worker, given the files that have already been sent to this worker and all decisions taken
previously; we select the best worker, hence the first \emph{min} in the heuristic. We take the
minimum of starting dates over all tasks, hence the second \emph{min}.
\end{description}

It turns out that neither greedy algorithm is optimal. See Figure~\ref{fig.contrex}
for an example where Min-min is better than Thrifty, and Figure~\ref{fig.contrex2}
for an example of the opposite situation.

\begin{figure}
    \subfigure[]{\includegraphics[height=40mm]{contrex.fig}\label{fig.contrex}}
    \hfill
    \subfigure[]{\includegraphics[height=40mm,subfig=1]{contrex2.fig}\label{fig.contrex2}}
    \caption{Neither Thrifty nor Min-min is optimal:
      (a) with $\nbp=2$, $\comm=4$, $\calc=7$, and $\rrr = \sss= 3$, Min-min wins;
      (b) with $\nbp=2$, $\comm=8$, $\calc=9$, $\rrr =6$, and $\sss= 3$, Thrifty wins.
    }
\end{figure}

We now go back to our original model.

\section{Minimization of the communication volume}
\label{sec.volume}

In this section, we derive a lower bound on the total number of
communications (sent from, or received by, the master) that are needed
to execute any matrix multiplication algorithm.  We point out that, since
we are not interested in optimizing the execution time (a difficult
problem, according to Section~\ref{sec.stripes}) but only in
minimizing the total communication volume, we can simulate any
parallel algorithm on a single worker. Therefore, we only need to
consider the one-worker case.

We deal with the original, and realistic, formulation of the
problem as follows:
\begin{itemize}
  \item The master sends blocks $\A_{ik}$, $\B_{kj}$, and $\C_{ij}$,
  \item The master retrieves final values of blocks $\C_{ij}$, and
  \item We enforce limited memory on the worker; only $\mem$ buffers
    are available, which means that at most $\mem$ blocks of $\A$,
    $\B$, and/or $\C$ can simultaneously be stored on the worker.
\end{itemize}

First, we describe an algorithm that aims at re-using $\C$ blocks as
much as possible after they have been loaded. Next, we assess the
performance of this algorithm. Finally, we improve a lower bound
previously established by Toledo~\cite{toledo99survey,toledoJPDC}.

\begin{figure}
  \begin{center}
    \includegraphics[width=\textwidth]{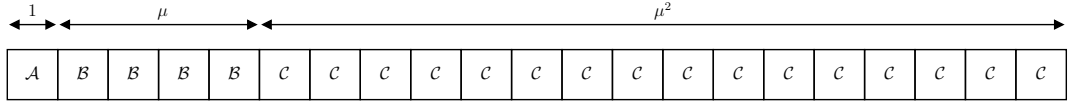}
  \end{center}
  \caption{Memory usage for the \emph{maximum re-use}
    algorithm when $\mem = 21$: $\mm = 4$; $1$ block is used for $\A$,
    $\mm$ for \B, and $\mm^2$ for \C.}
  \label{fig.maxproduct}
\end{figure}

\begin{figure}
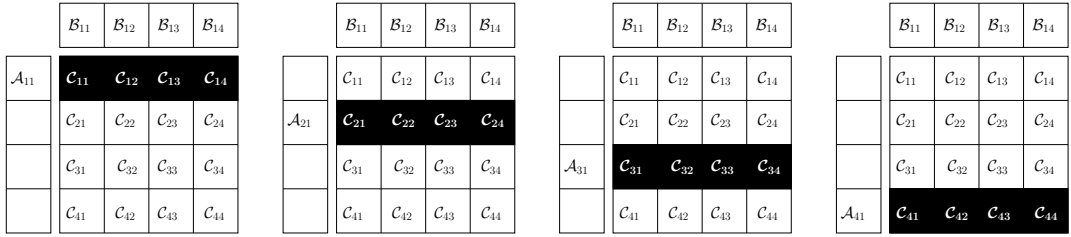

  \centering
  \begin{minipage}[b]{0.22\linewidth}
    \includegraphics[width=\textwidth,subfig=7]{MMgreedy.fig}
  \end{minipage}
  \hfill
  \begin{minipage}[b]{0.22\linewidth}
    \includegraphics[width=\textwidth,subfig=8]{MMgreedy.fig}
  \end{minipage}
  \hfill
  \begin{minipage}[b]{0.22\linewidth}
    \includegraphics[width=\textwidth,subfig=9]{MMgreedy.fig}
  \end{minipage}
  \hfill
  \begin{minipage}[b]{0.22\linewidth}
    \includegraphics[width=\textwidth,subfig=10]{MMgreedy.fig}
  \end{minipage}
  \caption{Four steps of the \emph{maximum re-use} algorithm, with $\mem = 21$ and $\mm = 4$. The elements of $\C$ updated are displayed on white on black.}
  \label{Mat.fig:maxreuse}
\end{figure}



\subsection{The \emph{maximum re-use} algorithm}

Below we introduce and analyze the performance of the \emph{maximum
  re-use} algorithm, whose memory management is illustrated in
Figure~\ref{fig.maxproduct}.  Four consecutive execution steps are
shown in Figure~\ref{Mat.fig:maxreuse}.  Assume that there are $\mem$
available buffers. First we find $\mm$ as the largest integer such
that $1 + \mm + \mm² \leq \mem$. The idea is to use one buffer to
store $\A$ blocks, $\mm$ buffers to store $\B$ blocks, and $\mm^2$
buffers to store $\C$ blocks. In the outer loop of the algorithm, a
$\mm \times \mm$ square of $\C$ blocks is loaded. Once these $\mm^2$
blocks have been loaded, they are repeatedly updated in the inner loop
of the algorithm until their final value is computed. Then the blocks
are returned to the master, and $\mm^2$ new $\C$ blocks are sent by
the master and stored by the worker. As illustrated in
Figure~\ref{fig.maxproduct}, we need $\mm$ buffers to store a row of
$\B$ blocks, but only one buffer for $\A$ blocks: $\A$ blocks are sent
in sequence, each of them is used in combination with a row of $\mm$
$\B$ blocks to update the corresponding row of $\C$ blocks. This leads
to the following sketch of the algorithm:

\noindent
\textbf{Outer loop}: while there remain $\C$ blocks to be computed
\begin{itemize}
  \item Store $\mm^2$  blocks of $\C$ in worker's memory:\\
  send a $\mm \times \mm$ square
  $\{\C_{i,j}\ /\ i_0 \leq i < i_0+\mm,\ j_0 \leq j < j_0+\mm \}$
  \item \textbf{Inner loop}: For each $k$ from $1$ to $\ttt$:
  \begin{enumerate}
  \item Send a row of $\mm$ elements $\{\B_{k,j}\ /\ j_0 \leq j <
  j_0+\mm \}$;
  \item Sequentially send $\mm$ elements of column $\{\A_{i,k}\ /\ i_0
  \leq i < i_0+\mm \}$. For each $A_{i,k}$, update $\mm$ elements of $\C$
  \end{enumerate}
  \item Return results to master.
\end{itemize}

\subsection{Performance and lower bound}

The performance of one iteration of the outer loop of the
\emph{maximum re-use} algorithm can readily be determined:
\begin{itemize}
  \item We need $2 \mm^2$ communications to send and retrieve $\C$ blocks.
  \item For each value of $\ttt$:\\
  - we need $\mm$ elements of $\A$ and $\mm$ elements of $\B$;\\
  - we update $\mm^2$ blocks.
\end{itemize}
In terms of block operations, the communication-to-computation ratio
achieved by the algorithm is thus
$$\CCRB = \frac{2 \mm^2 + 2 \mm \ttt}{\mm^2 \ttt} = \frac{2}{\ttt} + \frac{2}{\mm}.$$
For large problems, i.e., large values of $\ttt$, we see that $\CCRB$
is asymptotically close to the value $\CCRB_{\infty} =
\frac{2}{\sqrt{\mem}}$.  We point out that, in terms of data elements,
the communication-to-computation ratio is divided by a factor $q$.
Indeed, a block consists of $q^2$ coefficients but an update requires
$q^3$ floating-point operations.

How can we assess the performance of the \emph{maximum re-use} algorithm? How good is the value of $\CCRB$?
To see this, we refine an analysis due to Toledo~\cite{toledo99survey}.
The idea is to estimate the number of computations made thanks to $\mem$
consecutive communication steps (again, the unit is a matrix block here).
We need some notations:
\begin{itemize}
\item We let \aold, \betaold, and $\cold$ be the number of buffers
  dedicated to $\A$, $\B$, and $\C$ at the beginning of the $\mem$
  communication steps;
\item We let $\anew$, $\bnew$, and $\cnew$ be the number of $\A$,
  $\B$, and $\C$ blocks sent by the master during the $\mem$
  communication steps;
\item Finally, we let $\csend$ be the number of $\C$ blocks returned
  to the master during these $\mem$ steps.
\end{itemize}

Obviously, the following equations must hold true:

\begin{equation*}
  \left\{
    \begin{array}{l}
      \aold + \betaold + \cold \leq \mem\\
      \anew + \bnew + \cnew + \csend = \mem
    \end{array}
  \right.
\end{equation*}

The following lemma is given in~\cite{toledo99survey}: consider any
algorithm that uses the standard way of multiplying matrices (this
excludes Strassen's or Winograd's algorithm~\cite{CLRbook}, for
instance).  If $N_A$ elements of $\A$, $N_B$ elements of $\B$ and
$N_C$ elements of $\C$ are accessed, then no more than $K$
computations can be done, where
$$K = \min \left\{ (N_A+N_B)\sqrt{N_C}, (N_A+N_C)\sqrt{N_B}, (N_B+N_C)\sqrt{N_A} \right\}.$$
To use this result here, we see that no more than $\aold + \anew$
blocks of $\A$ are accessed, hence $N_A = (\aold + \anew)q^2$.
Similarly, $N_B = (\betaold + \bnew)q^2$ and $N_C = (\cold +
\cnew)q^2$ (the $\C$ blocks returned are already counted).  We
simplify notations by writing:
\begin{equation*}
\left\{
\begin{array}{l}
  \aold + \anew = \alpha \mem\\
  \betaold + \bnew = \beta \mem\\
  \cold + \cnew = \gamma \mem\\
\end{array}
\right.
\end{equation*}
Then we obtain
$$K = \min \left\{(\alpha + \beta)\sqrt{\gamma}, (\beta +
  \gamma)\sqrt{\alpha}, (\gamma + \alpha)\sqrt{\beta}\right\} \times
\mem \sqrt{\mem} q^3.$$
Writing $K = k \mem \sqrt{\mem} q^3$, we obtain the following system
of equations
\begin{equation*}
\left\{
\begin{array}{l}
  \textsc{Maximize~} k \textsc{~s.t.}\\
  \begin{split}
    k \leq (\alpha + \beta)\sqrt{\gamma}\\
    k \leq (\beta + \gamma)\sqrt{\alpha}\\
    k \leq (\gamma + \alpha)\sqrt{\beta}\\
    \alpha + \beta + \gamma \leq 2\\
  \end{split}\\
\end{array}
\right.
\end{equation*}
whose solution is easily found to be
$$\alpha = \beta = \gamma = \frac{2}{3}, \textsc{~and~} k = \sqrt{\frac{32}{27}}.$$
This gives a lower bound for the communication-to-computation ratio
(in terms of blocks) of any algorithm:
$$\CCRB_{\text{opt}} = \frac{\mem}{k \mem \sqrt{\mem}} = \sqrt{\frac{27}{32 \mem}}.$$

In fact, it is possible to refine this bound. Instead of using the lemma given in~\cite{toledo99survey},
we use Loomis-Whitney inequality~\cite{toledoJPDC}: if $N_A$ elements of $\A$, $N_B$ elements of $\B$, and
$N_C$ elements of $\C$ are accessed, then no more than $K$ computations can be done, where
$$K = \sqrt{N_A N_B N_C}.$$
Here
$$K = \sqrt{\alpha + \beta + \gamma} \times \mem
\sqrt{\mem} q^3$$
We obtain
$$\alpha = \beta = \gamma = \frac{2}{3}, \text{~and~} k = \sqrt{\frac{8}{27}},$$
so that the lower bound for the communication-to-computation ratio becomes:
$$\CCRB_{\text{opt}} = \sqrt{\frac{27}{8 \mem}}.$$
The \emph{maximum re-use} algorithm does not achieve the lower bound:
$$\CCRB_{\infty} = \frac{2}{\sqrt{\mem}} = \sqrt{\frac{32}{8 \mem}}$$
but it is quite close!

Finally, we point out that the bound $\CCRB_{\text{opt}}$ improves upon the best-known value
$\sqrt{\frac{1}{8 \mem}}$ derived in~\cite{toledoJPDC}. Also, the ratio $\CCRB_{\infty}$
achieved by the \emph{maximum re-use} algorithm is lower by a factor $\sqrt{3}$ than the ratio
achieved by the \emph{blocked matrix-multiply} algorithm of~\cite{toledo99survey}.

\section{Algorithms for homogeneous platforms}
\label{sec.homo}

In this section, we adapt the \emph{maximum re-use} algorithm to fully homogeneous platforms.
In this framework, contrary to the simplest
version, we have a limitation of the memory capacity. So we must
first decide which part of the memory will be used to stock which
part of the original matrices, in order to maximize the total number of
computations per time unit.
Cannon's algorithm~\cite{Cannon69} and the ScaLAPACK outer
product algorithm~\cite{Scalapack97} both distribute square blocks of $\C$ to the processors.
Intuitively, squares are better than elongated rectangles because their perimeter
(which is proportional to the communication volume) is smaller
for the same area. We use the same approach here, but we have not been able to assess any
optimal result.

\subsubsection*{Principle of the algorithm}

We load into the memory of each worker $\mm$ $q\times q$ blocks of $\A$ and $\mm$
$q\times q$ blocks of $\B$ to compute $\mm^2 \ q\times q$ blocks of
$\C$. In addition, we need $2\mm$ extra buffers, split into $\mm$ buffers for $\A$ and
$\mm$ for $\B$,  in order to overlap computation and communication
steps. In fact, $\mm$ buffers for $\A$ and $\mm$ for $\B$ would suffice for
each update, but we need to prepare for the next update while computing.
Overall, the number of $\C$ blocks that we can simultaneously load into memory is
the largest integer $\mm$ such that
$$\mm^2 + 4\mm \leq \mem.$$

We have to determine the number of participating workers $\usedp$. For
that purpose, we proceed as follows.  On the communication side, we
know that in a round (computing a $\C$ block entirely), the master
exchanges with each worker $2 \mm^2$ blocks of $\C$ ($\mm²$ sent and
$\mm²$ received), and sends $\mm \ttt $ blocks of $\A$ and $\mm \ttt$
blocks of $\B$. Also during this round, on the computation side, each
worker computes $\mm^2 \ttt$ block updates.

If we enroll too many processors, the communication capacity of the
master will be exceeded. There is a limit on the number of blocks sent
per time unit, hence on the maximal processor number $\usedp$, which
we compute as follows: $\usedp$ is the smallest integer such that
$$2 \mm \ttt \comm \times \usedp \geq \mm^2 \ttt \calc.$$
Indeed, this is the smallest value to saturate the communication
capacity of the master required to sustain the corresponding
computations. We derive that
$$\usedp  = \left\lceil \frac{\mm^2 \ttt \calc}{2 \mm \ttt\comm} \right\rceil
= \left\lceil \frac{\mm\calc}{2\comm} \right\rceil.$$

In the context of matrix multiplication, we have $\comm = q^2 \tau_c$
and $\calc = q^3 \tau_a$, hence $\usedp = \left\lceil \frac{\mm q}{2}
  \frac{\tau_a}{\tau_c} \right\rceil$.  Moreover, we need to enforce
that $\usedp \leq \nbp$, hence we finally obtain the formula
$$\usedp = \min \left \{
  \nbp, \left\lceil \frac{\mm q}{2} \frac{\tau_a}{\tau_c} \right\rceil \right
\}.$$

For the sake of simplicity, we suppose that $r$ is divisible by $\mm$,
and that $s$ is divisible by $\usedp \mm$.  We allocate $\mm$ block
columns (i.e., $q \mm$ consecutive columns of the original matrix) of
$\C$ to each processor. The algorithm is decomposed into two parts.
Algorithm~\ref{algo:homo_master} outlines the program of the master,
while Algorithm~\ref{algo:homo_slaves} is the program of each worker.

\begin{algorithm}[Hbt]
\label{algo:homo_master}
$\mm \la \left\lfloor \sqrt{4 + \mem} - 2 \right\rfloor$\; $\usedp \la
\min \Big \{ \nbp,\left\lceil \frac{\mm \calc}{2 \comm} \right\rceil
\Big \}$\;

Split the matrix into squares $\block_{i',j'}$ of $\mm^2$ blocks
(of size $q \times q$):\\
$\block_{i',j'} = \{\C_{i,j}\ \backslash\ (i'-1)\mm+1
\leq i \leq i' \mm, (j'-1)\mm+1 \leq j
\leq j'\mm \}$\;
\For {$j'' \la 0$ \KwTo $\frac{s}{\usedp\mm} \text{ by Step } \usedp$ } {
  \For {$i' \la 1$ \KwTo $\frac{r}{\mm}$} {
    \For{$id_{worker} \la  1$ \KwTo $\usedp$} {
      $j' \la j'' + id_{worker}$\;
      Send block $\block_{i', j'}$ to worker $id_{worker}$\;
    }
    \For{$k \la 1$ \KwTo $t$} {
      \For{$id_{worker} \la  1$ \KwTo $\usedp$} {
    $j' \la j'' + id_{worker}$\;
    \For{$j \la (j'-1)\mm+1$ \KwTo $j' \mm$}{
      Send $\B_{k,j}$\;
    }
    \For{$i \la (i'-1)\mm+1$ \KwTo $i'\mm$}{
      Send $\A_{i,k}$\;
    }
      }
    }
    \For{$id_{worker} \la  1$ \KwTo $\usedp$}{
      $j' \la j'' + id_{worker}$\;
      Receive $\block_{i',j'}$ from worker $id_{worker}$\;
    }
  }
}
\caption{Homogeneous version, master program.}
\end{algorithm}

\begin{algorithm}[Hbt]
\label{algo:homo_slaves}
\For{all blocks} {
  Receive $\block_{i',j'}$ from master\;
  \For{$k \la 1$ \KwTo $t$} {
    \lFor{$j \la (j'-1)\mm+1$ \KwTo $j'\mm$}{
      Receive $\B_{k,j}$\;
    }
    \For{$i \la (i'-1)\mm+1$ \KwTo $i'\mm$} {
      Receive $\A_{i,k}$\;
      \For{$j \la (j'-1)\mm+1$ \KwTo $j'\mm$}{
       $\C_{i,j} \la \C_{i,j} + \A_{i,k}.\B_{k,j}$\;
      }
    }
  }
  Return $\block_{i',j'}$ to master\;
}
\caption{Homogeneous version, worker program.}
\end{algorithm}

\subsubsection*{Impact of the start-up overhead}

If we follow the execution of the homogeneous algorithm, we may wonder
whether we can really neglect the input/output of $\C$ blocks.
Contrary to the greedy algorithms for the simplest instance
described in Section~\ref{sec.stripes}, we sequentialize here the
sending, computing, and receiving of the $\C$ blocks, so that each
worker loses $2 \comm$ time-units per block, i.e., per $\ttt \calc$
time-units. As there are $\usedp \leq \frac{\mm \calc}{2 \comm} + 1$
workers, the total loss would be of $2 \comm \usedp$ time-units every
$\ttt \calc$ time-units, which is less than $\frac{\mm}{\ttt} +
\frac{2\comm}{\ttt\calc}$.  For example, with $\comm = 2$, $\calc =
4.5$, $\mm=4$ and $\ttt=100$, we enroll $\usedp = 5$ workers, and the
total lost is at most $4\%$, which is small enough to be neglected.
Note that it would be technically possible to design an algorithm
where the sending of the next block is overlapped with the last
computations of the current block, but the whole procedure gets much
more complicated.

\subsubsection*{Dealing with ``small'' matrices or platforms}

We have shown that our algorithm should use $\usedp = \min \left \{
  \nbp, \left\lceil \frac{\mm q}{2} \frac{\tau_a}{\tau_c} \right\rceil
\right \}$ processors, each of them holding $\mm^2$ blocks of matrix
$\mathcal{C}$. For this solution to be feasible, $\mathcal{C}$ must be
large enough. In other words, this solution can be implemented if and
only if $\rrr \times \sss \geq \min \left \{ \nbp, \left\lceil \frac{\mm
      q}{2} \frac{\tau_a}{\tau_c} \right\rceil \right \} \mm^2$. If
$\mathcal{C}$ is not large enough, we will only use $\usedq < \usedp$
processors, each of them holding $\nu^2$ blocks of $\mathcal{C}$, such
that:
$$
\left\{
\begin{array}{l}
  \usedq \nu^2 \leq \rrr \times \sss\\
  \nu^2 \calc \leq 2 \nu \usedq \comm
\end{array}
\right.
\qquad
\Leftrightarrow
\qquad
\left\{
\begin{array}{l}
  \usedq \nu^2 \leq \rrr \times \sss\\
  \frac{\nu \calc}{2 \comm} \leq  \usedq
\end{array}
\right.,
$$
following the same line of reasoning as previously. We obviously
want $\nu$ to be the largest possible in order for the communications
to be most beneficial. For a given value of $\nu$ we want $\usedq$ to
be the smallest to spare resources. Therefore, the best solution is
given by the largest value of $\nu$ such that:
$$\left\lceil\frac{\nu \calc}{2 \comm}\right\rceil \nu^2 \leq \rrr \times \sss,$$
and then $\usedq = \left\lceil\frac{\nu \calc}{2 \comm}\right\rceil$.

If the platform does not contain the desired number of processors,
i.e., if $\usedp > \nbp$ in the case of a ``large'' matrix
$\mathcal{C}$ or if $\usedq > \nbp$ otherwise, then we enroll all the
$\nbp$ processors and we give them $\nu^2$ blocks of $\mathcal{C}$
with $\nu
=\min\left\{\frac{\rrr\times\sss}{\nbp},\frac{2\comm}{\calc}\nbp\right\}$,
following the same line of reasoning as previously.

\section{Algorithms for heterogeneous platforms}
\label{sec.hetero}

In this section, all processors are heterogeneous, in term of memory
size as well as computation or communication time.  As in the previous
section, $\mem_i$ is the number of $q \times q$ blocks that fit in the
memory of worker $P_i$, and we need to load into the memory of $P_i$
$2\mm_i$ blocks of $\A$, $2\mm_i$ blocks of $\B$, and $\mm_i^2$ blocks
of $\C$.  This number of blocks loaded into the memory changes from
worker to worker, because it depends upon their memory capacities. We
first compute all the different values of $\mm_i$ so that
$$\mm_i^2 + 4\mm_i \leq \mem_i.$$

To adapt our \emph{maximum re-use} algorithm to heterogeneous
platforms, we first design a greedy algorithm for resource selection
(Section~\ref{sec.badgreedy}), and we discuss its limitations. We
introduce our final algorithm for heterogeneous platforms in
Section~\ref{sec.hetfinal}.

\subsection{Bandwidth-centric resource selection}
\label{sec.badgreedy}

Each worker $P_i$ has parameters $\comm_i$, $\calc_i$, and $\mm_i$,
and each participating $P_i$ needs to receive $\delta_i = 2 \mm_i \ttt
\comm_i$ blocks to perform $\phi_i = \ttt \mm_i^2 \calc_i$
computations. Once again, we neglect I/O for $\C$ blocks.  Consider
the steady-state of a schedule. During one time-unit, $P_i$ receives a
certain amount $y_i$ of blocks, both of $\mathcal{A}$ and
$\mathcal{B}$, and computes $x_i$ $\C$ blocks. We express the
constraints, in terms of communication ---the master has limited
bandwidth--- and of computation ---a worker cannot perform more work
than it receives.  The objective is to maximize the amount of work
performed per time-unit. Altogether, we gather the following linear
program:
\begin{center}
  \begin{equation*}
    \left\{
    \begin{array}{r}
      \textsc{Maximize~} \sum_i x_i\\
      \textsc{subject to}\\
      \begin{split}
        ~~~~\sum_i y_i \comm_i \leq 1\\
        \forall i,~~x_i \calc_i \leq 1\\
        \forall i,~~\frac{x_i}{\mm_i^2} \leq \frac{y_i}{2 \mm_i}\\
      \end{split}\\
    \end{array}
    \right.
  \end{equation*}
\end{center}
Obviously, the best solution for $y_i$ is $y_i = \frac{2 x_i}{\mm_i}$,
so the problem can be reduced to :
$$
\begin{array}{l}
  \left\{
  \begin{array}{l}
    \textsc{Maximize~} \sum_i x_i\\
    \textsc{subject to}\\
      \forall i,~~x_i \leq \frac{1}{\calc_i}\\
      ~~~~\sum_i \frac{2 \comm_i}{\mm_i} x_i \leq 1\\
  \end{array}
  \right.
\end{array}
$$
The optimal solution for this system is a bandwidth-centric
strategy~\cite{c122,j87}; we sort workers by
non-decreasing values of $\frac{2 \comm_i}{\mm_i}$ and we enroll them as long
as $\sum \frac{2 \comm_i}{\mm_i \calc_i} \leq 1$. In this way, we can
achieve the throughput $\rho \approx \sum_{i \text{~enrolled}}
\frac{1}{\calc_i}$.

\vspace{\baselineskip}

This solution seems to be close to the optimal. However, the problem
is that workers may not have enough memory to execute it!  Consider
the example described by Table~\ref{tab.bandwidthcentric}.

\begin{table}
  \begin{minipage}[b]{.48\linewidth}
    \begin{center}
      \begin{tabular}{|c|c|c|}
        \hline
        & $P_1$ & $P_2$ \\\hline
        $\comm_i$   & $1$     & $20$ \\\hline
        $\calc_i$   & $2$     & $40$  \\\hline
        $\mm_i$ & $2$     & $2$ \\\hline
        $\frac{2c_i}{\mm_i \calc_i}$ & $\frac{1}{2}$ & $\frac{1}{2}$ \\[1mm]\hline
      \end{tabular}
    \end{center}
    \caption{Platform for which the bandwidth centric solution is not feasible.\label{tab.bandwidthcentric}}
  \end{minipage}
  \hfill
  \begin{minipage}[b]{.48\linewidth}
    \begin{center}
      \begin{tabular}{|c|c|c|c|}
        \hline
        & $P_1$ & $P_2$ & $P_3$\\\hline
        $\comm_i$   & 2     & 3     & 5 \\\hline
        $\calc_i$   & 2     & 3     & 1 \\\hline
        $\mm_i$ & 6     & 18    & 10 \\\hline
        $\mm_i^2$ & 36 & 324 & 100 \\\hline
        $2 \mm_i \comm_i$ & 24 & 108 & 100\\\hline
      \end{tabular}
    \end{center}
    \caption{Platform used to demonstrate the processor selection algorithms.\label{tab.littleex}}
  \end{minipage}
\end{table}


Using the bandwidth-centric strategy, every $160$ seconds:
\begin{itemize}
\item $P_1$ receives $80$ blocks ($20$ $\mm_1 \times \mm_1$ chunks) in
  $80$ seconds, and computes $80$ blocks in $160$ seconds;
\item $P_2$ receives $4$ blocks ($1$ $\mm_2 \times \mm_2$ chunk) in
  $80$ seconds, and computes $4$ blocks in $160$ seconds.
\end{itemize}
But $P_1$ computes two quickly, and it needs buffers to store as many
as $20$ blocks to stay busy while one block is sent to $P_2$:
$$\begin{array}{cccccc}
Communications & 11111111111111111111 & 20  & 11111111111111111111 & 20 & 111111111 \ldots\\
Processor & P_1 & P_2 & P_1 & P_2 & P_1 \ldots
\end{array}$$

Therefore, the bandwidth-centric solution cannot always be realized in
practice, and we turn to another algorithm described below. To avoid
the previous buffer problems, resource selection will be performed
through a step-by-step simulation. However, we point out that the
steady-state solution can be seen as an upper bound of the performance
that can be achieved.

\subsection{Incremental resource selection}
\label{sec.hetfinal}

The different memory capacities of the workers imply that we assign them chunks
of different sizes. This requirement complicates the global partitioning of the $\C$ matrix
among the workers. To take this into account and simplify the implementation, we decide
to assign only full matrix column blocks in the algorithm. This is done in a two-phase
approach.

In the first phase we pre-compute the allocation of blocks to
processors, using a processor selection algorithm we will describe
later. We start as if we had a huge matrix of size $\infty \times
\sum_{i=1}^{n} \mm_i$. Each time a processor $P_i$ is chosen by the
processor selection algorithm it is assigned a square chunk of
$\mm_i^2$ $\C$ blocks. As soon as some processor $P_i$ has enough
blocks to fill up $\mm_i$ block columns of the initial matrix, we
decide that $P_i$ will indeed execute these columns during the
parallel execution. Therefore we maintain a panel of $\sum_{i=1}^{p}
\mm_i$ block columns and fill them out by assigning blocks to
processors.  We stop this phase as soon as all the $\rrr \times \sss$
blocks of the initial matrix have been allocated columnwise by this
process.  Note that worker $P_i$ will be assigned a block column after
it has been selected $\lceil \frac{\rrr}{\mm_i} \rceil$ times by the
algorithm.

In the second phase we perform the actual execution. Messages will be
sent to workers according to the previous selection process. The first
time a processor $P_i$ is selected, it receives a square chunk of
$\mm_i^2$ $\C$ blocks, which initializes its repeated pattern of
operation: the following $\ttt$ times, $P_i$ receives $\mm_i$ $\A$ and
$\mm_i$ $\B$ blocks, which requires $2 \mm_i \comm_i$ time-units.

There remains to decide which processor to select at each step.
We have no closed-form formula for the allocation of blocks to processors. Instead, we
use an incremental algorithm to compute which worker the next blocks will be assigned to.
We have two variants of the incremental algorithm, a \emph{global} one that aims at optimizing
the overall communication-to-computation ratio, and a \emph{local} one that selects the best processor
for the next stage. Both variants are described below.

\subsubsection{Global selection algorithm}

The intuitive idea for this algorithm is to select the processor that
maximizes the ratio of the total work achieved so far (in terms of
block updates) over the completion time of the last communication. The
latter represents the time spent by the master so far, either sending
data to workers or staying idle, waiting for the workers to finish
their current computations. We have:

$$\ratio \la \frac{\text{total work achieved}}{\text{completion time of last communication}}$$

Estimating computations is easy: $P_i$ executes $\mm_i^2$ block
updates per assignment. Communications are slightly more complicated
to deal with; we cannot just use the communication time $2 \mm_i
\comm_i$ of $P_i$ for the $\mathcal{A}$ and $\mathcal{B}$ blocks because we
need to take its ready time into account. Indeed, if $P_i$ is
currently busy executing work, it cannot receive additional data too
much in advance because its memory is limited.
Algorithm~\ref{alg:global} presents this selection process, which we iterate
until all blocks of the initial matrix are assigned and computed.



\begin{algorithm}[Htb]
\KwData{\\
$\complet$: the completion time of the last communication\\
$\ready_i$: the completion time of the work assigned to processor $P_i$\\
$\nblock_i$: the number of $A$ and $B$ blocks sent to processor $P_i$\\
$\totwork$: the total work assigned so far (in terms of block updates)\\
$\nbcol$: the number of fully processed  $\mathcal{C}$ block columns
}
\BlankLine
\textbf{INITIALIZATION}
\BlankLine
$\complet \la 0$\;
$\totwork \la 0$\;
\For{$i \la 1$ \KwTo $\nbp$}{%
  $\ready_i \la 0$\;
  $\nblock_i \la 0$\;
}
\BlankLine
\textbf{SIMULATION}
\BlankLine
\Repeat{$\nbcol \geq \sss$}{
  $\nextt \la$ worker that realizes
  $\max_{i=1}^{\nbp} \frac{\totwork + \mm_i^2}{\max(\complet + 2 \mm_i \comm_i,
    \ready_i)}$\;
  $\totwork \la \totwork + \mm_{\nextt}^2$\;
  $\complet \la \max(\complet + 2 \mm_{\nextt} \comm_{\nextt}, \ready_{\nextt})$\;
  $\ready_{\nextt} \la \complet + \mm_{\nextt}^2 \calc_{\nextt}$\;
  $\nblock_{\nextt} \la \nblock_{\nextt} + 2\mm_{\nextt}$\;
  $\nbcol \la \sum_{i=1}^{\nbp} \left\lfloor\frac{\nblock_i}{2\mm_i t \lceil
  \frac{\rrr}{\mm_i} \rceil} \right\rfloor \mm_i$\;
}
\caption{Global selection algorithm.\label{alg:global}}
\end{algorithm}


\paragraph{Running the global selection algorithm on an example.}

Consider the example described in Table~\ref{tab.littleex} with three
workers $P_1$, $P_2$ and $P_3$. For the first step, we have $\ratio_i
\la \frac{\mm_i^2}{2 \mm_i \comm_i}$ for all $i$.  We compute
$\ratio_1 = 1.5$, $\ratio_2 = 3$, and $\ratio_3 = 1$ and select $P_2$:
$\nextt \la 2$.  We update variables as $\totwork \la 0 + 324 = 324$,
$\complet \la \max(0 + 108, 0) = 108$, $\ready_{2} \la 108 + 972 =
1080$ and $\nblock_2 \la 36$.

At the second step we compute $\ratio_1 \la \frac{324+36}{108+24} =
2.71$, $\ratio_2 \la \frac{324+324}{1080} = 0.6$ and $\ratio_3 \la
\frac{324+100}{108+100} = 2.04$ and we select $P_1$. We point out that
$P_2$ is busy until time $t=1080$ because of the first assignment,
which we correctly took into account when computing $\ready_2$. For
$P_1$ and $P_3$ the communication could take place immediately after
the first one.  There remains to update variables: $\totwork \la 324 +
36 = 360$, $\complet \la \max(108+24, 0) = 132$, $\ready_{1} \la
132 + 72 = 204$ and $\nblock_1 \la 12$.

\begin{figure}[htb]
      \centering
      \includegraphics[width=.8\textwidth,subfig=13]{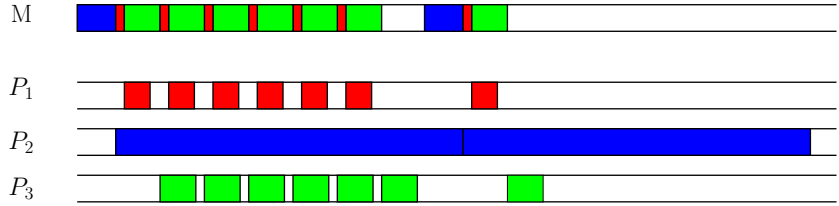}
      \caption{Global selection algorithm on the example of Table~\ref{tab.littleex}.}
      \label{fig.global-algo}
\end{figure}

At the third step the algorithm selects $P_3$. Going forward, we have a
cyclic pattern repeating, with $13$ consecutive communications, one to
$P_2$ followed by $12$ ones alternating between $P_1$ and $P_3$, and
then some idle time before the next pattern (see
Figure~\ref{fig.global-algo}). The asymptotic value of $\ratio$ is
$1.17$ while the steady-state approach of Section~\ref{sec.badgreedy}
would achieve a ratio of $1.39$ without memory limitations.  Finally,
we point out that it is easy to further refine the algorithm to get
closer to the performance of the steady-state. For instance, instead of
selecting the best processor greedily, we could look two-steps ahead
and search for the best pair of workers to select for the next two
communications (the only price to pay is an increase in the cost of
the selection algorithm).  From the example, the two-step ahead strategy
achieves a ratio $1.30$.

\subsubsection{Local selection algorithm}

The global selection algorithm picks, as the next processor, the one
that maximizes the ratio of the total amount of work
assigned over the time needed to send all the required data. Instead,
the local selection algorithm chooses, as destination of the $i$-th
communication, the processor that maximizes the ratio of the amount of
work assigned by this communication over the time during which the communication link
is used to performed this communication (i.e., the elapsed time
between the end of $(i-1)$-th communication and the end of the $i$-th
communication). As previously, if processor $P_j$ is the target of the
$i$-th communication, the $i$-th communication is the sending of
$\mm_j$ blocks of $\mathcal{A}$ and $\mm_j$ blocks of $\mathcal{B}$ to
processor $P_j$, which enables it to perform $\mm_j^2$ updates.

More formally, the local selection algorithm picks the worker $P_i$
that maximizes:
$${\frac{\mm_i²}{\max \{2\mm_i\comm_i,\ \ready_i-\complet\}}}$$

\begin{figure}[htb]
      \centering
      \includegraphics[width=.8\textwidth,subfig=14]{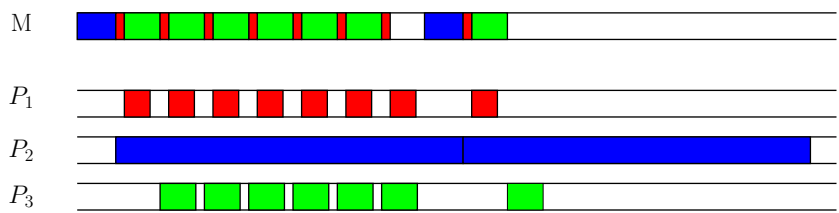}
      \caption{Local selection algorithm on the example of Table~\ref{tab.littleex}.}
      \label{fig.local-algo}
\end{figure}

Once again we consider the example described in
Table~\ref{tab.littleex}. For the first three steps, the global and
selection algorithms make the same decision. In fact, they take the
same first 13 decisions. However, for the 14-th selection, the global
algorithm picks processor $P_2$ when the local selection selects
processor $P_1$ and then processor $P_2$ for the 15-th decision, as
illustrated in Figure~\ref{fig.local-algo}. Under both selection
processes, the second chunk of work is sent to processor $P_2$ at the
same time but the local algorithm inserts an extra
communication. For this example, the local selection algorithm
achieves an asymptotic ratio of computation per communication of 1.21.
This is better than what is achieved by the global selection algorithm
but, obviously, there are examples where the global selection will
beat the local one.

\section{Extension to LU factorization}
\label{sec.LU}

In this section, we show how our techniques can be extended to LU
factorization. We first consider (Section~\ref{sec.lu.mono}) the case of a single worker,
in order to study how we can minimize the communication
volume. Then we present algorithms for
homogeneous clusters (Section~\ref{sec.lu.homo}) and for heterogeneous
platforms (Section~\ref{sec.lu.hetero}).

We consider the right-looking version of the LU factorization as it is
more amenable to parallelism. As previously, we use a block-oriented
approach. The atomic elements that we manipulate are not matrix
coefficients but instead square \emph{blocks} of size $q \times q$
(hence with $q^2$ coefficients). The size of the matrix is then $\rrr
\times \rrr$ blocks. Furthermore, we consider a second level of
blocking of size $\mm$. As previously, $\mm$ is the largest integer
such that $\mm^2 + 4\mm \leq \mem$. The main kernel is then a
rank-$\mm$ update $C \la C + A.B$ of blocks.  Hence the similarity
between matrix multiplication and LU decomposition.

\subsection{Single processor case}
\label{sec.lu.mono}

\begin{figure}
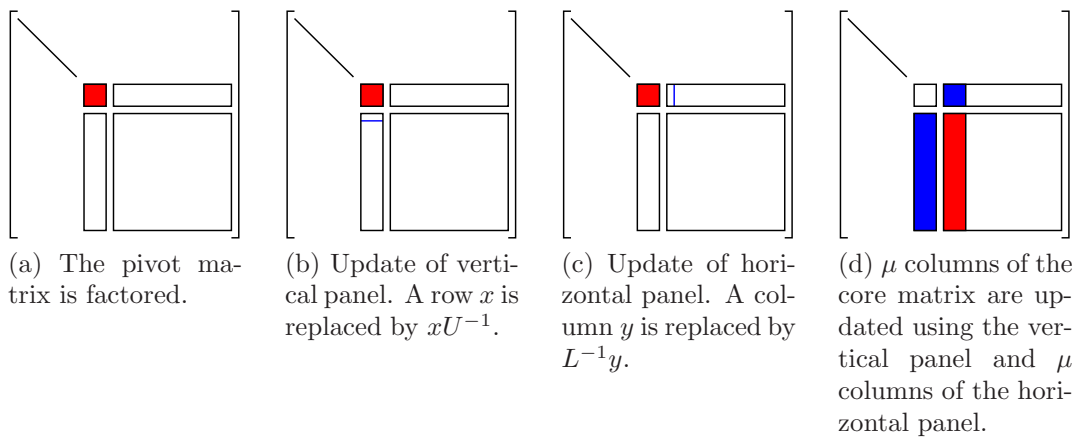

  \centering
  \begin{minipage}[t]{0.22\linewidth}
    \includegraphics[width=\textwidth,subfig=2]{LU.fig}
    (a) The pivot matrix is factored.
  \end{minipage}
  \hfill
  \begin{minipage}[t]{0.22\linewidth}
    \includegraphics[width=\textwidth,subfig=3]{LU.fig}
    (b) Update of vertical panel. A row $x$ is replaced
    by $xU^{-1}$.
  \end{minipage}
  \hfill
  \begin{minipage}[t]{0.22\linewidth}
    \includegraphics[width=\textwidth,subfig=10]{LU.fig}
    (c) Update of horizontal panel. A column $y$ is replaced
    by $L^{-1}y$.
  \end{minipage}
  \hfill
  \begin{minipage}[t]{0.22\linewidth}
    \includegraphics[width=\textwidth,subfig=11]{LU.fig}
    (d) $\mm$ columns of the core matrix are updated using the
    vertical panel and $\mm$ columns of the horizontal panel.
  \end{minipage}
  \caption{Scheme for LU factorization at step $k$.}
  \label{fig.LU}
\end{figure}

The different steps of LU factorization are presented in
Figure~\ref{fig.LU}.  Step $k$ of
the factorization consists of the following:
\begin{enumerate}
\item Factor pivot matrix (Figure~\ref{fig.LU}(a)). We compute at each step a pivot matrix of
  size $\mm^2$ (which thus contains $\mm^2 \times q^2$ coefficients). This
  factorization has a communication cost of $2\mm^2\comm$ (to bring the
  matrix and send it back after the update) and a computation cost of
  $\mm^3\calc$.

\item Update the $\mm$ columns below the pivot matrix (vertical panel)
  (Figure~\ref{fig.LU}(b)). Each row $x$ of this vertical panel is of
  size $\mm$ and must be replaced by $xU^{-1}$ for a computation cost
  of $\frac{1}{2}\mm^2 \calc$.

  The most communication-efficient policy to implement this update is
  to keep the pivot matrix in place and to move around the rows of the
  vertical panel.  Each row must be brought and sent back after
  update, for a total communication cost of $2\mm\comm$.

  At the $k$-th step, this update has then an overall communication
  cost of $2\mm(\rrr-k\mm)\comm$ and an overall computation cost
  of $\frac{1}{2}\mm^2(\rrr-k\mm)\calc$.

\item Update the $\mm$ rows at the right of the pivot matrix
  (horizontal panel) (Figure~\ref{fig.LU}(c)). Each column $y$ of this
  horizontal panel is of size $\mm$ and must be replaced by $L^{-1}y$
  for a computation cost of $\frac{1}{2}\mm^2 \calc$.

  This case is symmetrical to the previous one. Therefore, we follow
  the same policy and at the $k$-th step, this update has an
  overall communication cost of $2\mm(\rrr-k\mm)\comm$ and an
  overall computation cost of $\frac{1}{2}\mm^2(\rrr-k\mm)\calc$.

\item Update the core matrix (square matrix of the last $(\rrr-k\mm)$
  rows and columns) (Figure~\ref{fig.LU}(d)). This is a rank-$\mm$
  update. Contrary to matrix multiplication, the most
  communication-efficient policy is to not keep the result
  matrix in memory, but either a $\mm \times \mm$ square block of the vertical
  panel or of the horizontal panel (both solutions are symmetrical).
  Arbitrarily, we then decide to keep in memory a chunk of the
  horizontal panel. Then to update a row vector $x$ of the core
  matrix, we need to bring to that vector the corresponding row of the
  vertical panel, and then to send back the updated value of $x$. This
  has a communication cost of $3\mm\comm$ and a computation cost of
  $\mm^2$.

  At the $k$-th step, this update for $\mm$ columns of the core matrix
  has an overall communication cost of $(\mm^2+3(\rrr-k\mm)\mm)\comm$
  (counting the communications necessary to initially bring the
  $\mm^2$ elements of the horizontal panel) and an overall computation
  cost of $(\rrr-k\mm)\mm^2\calc$.

  Therefore, at the $k$-th step, this update has an overall
  communication cost of $(\frac{\rrr}{\mm}-k)(\mm^2+3(\rrr-k\mm)\mm)\comm$
  and an overall computation cost of
  $(\frac{\rrr}{\mm}-k)(\rrr-k\mm)\mm^2\calc$.
\end{enumerate}
Using the above scheme, the overall communication cost of the LU
factorization is
$$
\sum_{k=1}^{\frac{\rrr}{\mm}} \left(2\mm^2 + 4\mm(\rrr-k\mm) + \left(\frac{\rrr}{\mm}-k\right)(\mm^2+3(\rrr-k\mm)\mm)\right)
\comm = \left(\frac{\rrr^3}{\mm} - \rrr^2 + 2\mm\rrr\right)\comm,
$$
while the overall computation cost is
$$
\sum_{k=1}^{\frac{\rrr}{\mm}} \left(\mm^3 + \mm^2(\rrr-k\mm) + \left(\frac{\rrr}{\mm}-k\right)(\rrr-k\mm)\mm^2\right)\calc=
\frac{1}{3}\left(\rrr^3 + 2\mm^2\rrr\right)\calc.
$$

\subsection{Algorithm for homogeneous clusters}
\label{sec.lu.homo}

The most time-consuming part of the factorization is the update of the
core matrix (it has an overall cost of $\left(\frac{1}{3}\rrr^3 -
  \frac{1}{2}\mm\rrr^2 +\frac{1}{6}\mm^2\rrr\right)\calc$). Therefore,
we want to parallelize this update by allocating blocks of $\mm$
columns of the core matrix to different processors. Just as for
matrix multiplication, we would like to determine the optimal number
of participating workers $\usedp$.
For that purpose, we proceed as previously.  On the communication
side, we know that in a round (each worker updating $\mm$ columns
entirely), the master sends to each worker $\mm^2$ blocks of the
horizontal panel, then sends to each worker the $\mm(\rrr-k\mm)$
blocks of the vertical panel, and exchanges with each of them
$2\mm(\rrr-k\mm)$ blocks of the core matrix ($\mm(\rrr-k\mm)$ received
and later sent back after update).  Also during this round, on the
computation side, each worker computes $\mm^2(\rrr-k\mm)$ block
updates.
If we enroll too many processors, the communication capacity of the
master will be exceeded. There is a limit on the number of blocks sent
per time unit, hence on the maximal processor number $\usedp$, which
we compute as follows: $\usedp$ is the smallest integer such that
$$(\mm^2 + 3\mm(\rrr-k\mm))\comm\usedp \geq \mm^2(\rrr-k\mm) \calc.$$
We obtain that
$$\usedp  = \left\lceil \frac{\mm\calc}{3\comm} \right\rceil,$$
while neglecting the term $\mm^2$ in the communication cost, as we
assume $\frac{\rrr}{\mm}$ to be large.

Once the resource selection is performed, we propose a straightforward
algorithm: a single processor is responsible for the factorization of
the pivot matrix and of the update of the vertical and horizontal
panels, and then $\usedp$ processors work in parallel at the update of
the core matrix.

\subsection{Algorithm for heterogeneous platforms}
\label{sec.lu.hetero}

In this section, we simply sketch the algorithm for heterogeneous platforms.
When targeting heterogeneous platforms, there is a big difference
between LU factorization and matrix multiplication. Indeed, for LU
once the size $\mm$ of the pivot matrix is fixed, all processors have
to deal with it, whatever their memory capacities. There was no such
fixed common constant for matrix multiplication. Therefore, a crucial
step for heterogeneous platforms is to determine the size $\mm$ of the
pivot matrix. Note that two pivot matrices at two different steps of
the factorization may have different sizes, the constraint is that
all workers must use the same size at any given step of the elimination.

In theory, the memory size of the workers can be arbitrary.
In practice however, memory size usually is an
integral number of Gigabytes, and at most a few tens of Gigabytes. So
it is feasible to exhaustively study all the possible values of $\mm$,
estimate the processing time for each value, and then pick the best
one. Therefore, in the following we assume the value of $\mm$ has been
chosen, i.e., the pivot matrix is of a known size $\mm\times \mm$.

The memory layout used by each slave $P_i$ follows the same
policy than as for the homogeneous case:\\
- a chunk of the horizontal panel
is kept in memory,\\
- rows of the horizontal panel are sent to $P_i$,\\
- and rows of the core matrix are sent to $P_i$ and are returned to the
master after update.

If $\mm_i = \mm$, processor $P_i$ operates
exactly as for the homogeneous case. But if the memory
capacity of $P_i$ does not perfectly correspond to the size chosen for
the pivot matrix, we still have to decide the shape of the chunk of
the horizontal panel that processor $P_i$ is going to keep in its memory.
We have two cases to consider:
\begin{enumerate}
\item $\mm_i < \mm$. In other words, $P_i$ has not enough memory. Then
  we can imagine two different shapes for the horizontal panel chunk:
  \begin{enumerate}
  \item Square chunk, i.e., the chunk is of size $\mm_i \times \mm_i$.
    Then, for each update the master must send to $P_i$ a row of size
    $\mm_i$ of the horizontal panel and a row of size $\mm_i$ of the
    core matrix, and $P_i$ sends back after update the row of the core
    matrix. Hence a communication cost of $3\mm_i\comm$ for $\mm_i^2$
    computations. The computation-to-communication cost induced by
    this chunk shape is then:
    $$\frac{\mm_i^2\calc}{3\mm_i\comm} = \frac{\mm_i\calc}{3\comm}.$$

  \item Set of whole columns of the horizontal panel, i.e., the chunk
    is of size $\mm \times \left(\frac{\mm_i^2}{\mm}\right)$.  Then,
    for each update the master must send to $P_i$ a row of size $\mm$ of
    the horizontal panel and a row of size $\frac{\mm_i^2}{\mm}$ of
    the core matrix, and $P_i$ sends back after update the row of the
    core matrix. Hence a communication cost of $\left(\mm +
    2\frac{\mm_i^2}{\mm}\right)\comm$ for $\mm_i^2$ computations. The
    computation to communication cost induced by this chunk shape is
    then:
    $$\frac{\mm_i^2\calc}{\left(\mm + 2\frac{\mm_i^2}{\mm}\right)\comm}.$$
  \end{enumerate}
  The choice of the policy depends on the ratio $\frac{\mm_i}{\mm}$.
  Indeed,
  $$
  \frac{\mm_i^2\calc}{3\mm_i\comm} < \frac{\mm_i^2\calc}{\left(\mm +
      2\frac{\mm_i^2}{\mm}\right)\comm}
  \quad \Leftrightarrow \quad
  {\left(\mm + 2\frac{\mm_i^2}{\mm}\right)\comm} < {3\mm_i\comm}
  \quad \Leftrightarrow \quad
  %
  %
  %
  \left(2\frac{\mm_i}{\mm}-1\right)\left(\frac{\mm_i}{\mm}-1\right) <
  0.$$
  Therefore, the square chunk approach is more efficient if and
  only if $\mm_i \leq \frac{1}{2} \mm$.
\item $\mm_i > \mm$. In other words, $P_i$ has more memory than
  necessary to hold a square matrix like the pivot matrix, that is a
  matrix of size $\mm\times\mm$. In that case, we propose to divide
  the memory of $P_i$ into
  $\left\lfloor\frac{\mm_i^2}{\mm^2}\right\rfloor$ square chunks of
  size $\mm$, and to use this processor as if there were in fact
  $\left\lfloor\frac{\mm_i^2}{\mm^2}\right\rfloor$ processors with a
  memory of size $\mm^2$.
\end{enumerate}

So far, we have assumed we knew the value of $\mm$ and we have
proposed memory layout for the workers. We still have to decide which
processor to enroll in the computation. We perform the resource
selection as for matrix multiplication: we decide to assign only full
matrix column blocks of the core matrix and of the horizontal panel to
workers, and we actually perform resource selection using the same
selection algorithms than for matrix-multiplication.

The overall process to define a solution is then:
\begin{enumerate}
\item For each possible value of $\mm$ do
  \begin{enumerate}
  \item Find the processor which will be the fastest to factor the
    pivot matrix,  and to update the horizontal and vertical panels.
  \item Perform resource selection and then estimate the running time
    of the update of the core-matrix.
  \end{enumerate}
\item Retain the solution leading to the best (estimated) overall
  running time.
\end{enumerate}

\section{MPI experiments}
\label{sec.MPI}

In this section, we aim at validating the previous theoretical
results and algorithms. We conduct a variety of MPI experiments
to compare our new schemes with several other algorithms from the literature.
In the final version of this paper, we will report results obtained
for heterogeneous platforms, assessing the impact of the degree of heterogeneity
(in processor speed, link bandwidth and memory capacity) on the
performance of the various algorithms.
For this current version, we restrict to homogeneous platforms. Even in this
simpler framework, using a sophisticated memory management turns
out to be very important.

We start with a description of the platform, and of all the different algorithms
that we  compare. Then we describe
the experiments that we have conducted and justify their purpose. Finally,
we discuss the results.

\subsection{Platform}

For our experiments we are using a platform at the University of
Tennessee. All experiments are performed on a cluster of 64 Xeon 3.2GHz dual-processor
nodes. Each node of the cluster has four Gigabytes of memory and runs the Linux operating
system. The nodes are connected with a switched 100Mbps Fast Ethernet network.
In order to build a master-worker platform, we arbitrarily choose one
processor as the master, and the other processors become the workers.
Finally we used \emph{MPI\_WTime} as timer in all experiments.

\subsection{Algorithms}

We choose six different algorithms
from the general literature to compare our algorithm to.
We partition these algorithms into two sets. The first
set is composed of algorithms which use the same memory allocation
than ours. The only difference
between the algorithms is the order in which the master sends blocks to workers.
\begin{description}
\item[Homogeneous algorithm] (\textbf{HoLM}) is our homogeneous algorithm. It makes resource
  selection, and sends blocks to the selected workers in a round-robin fashion.
\item[Overlapped Round-Robin, Optimized Memory Layout] (\textbf{ORROML}) is very
  similar to our homogeneous algorithm. The only difference between
  them is that it does not make any resource selection, and so sends
  tasks to all available workers in a round-robin fashion.
\item[Overlapped Min-Min, Optimized Memory Layout] (\textbf{OMMOML}) is
  a static scheduling heuristic, which sends the next block to the first
  worker that will be available to compute it.
  As it is looking for
  potential workers in a given order, this algorithm performs some resource
  selection too. Theoretically, as our homogeneous resource selection
  ensures that the first worker is free to compute when we finish to
  send blocks to the others, they should have similar behavior.
\item[Overlapped Demand-Driven, Optimized Memory Layout]
  (\textbf{ODDOML}) is a demand-driven algorithm. In order to use the extra buffers available in the worker
  memories, it will send the next block to the
  first worker which can receive it. This would be a dynamic
  version of our algorithm, if it took
  worker selection into account.
\item[Demand-Driven, Optimized Memory Layout] (\textbf{DDOML}) is a very
  simple dynamic demand-driven algorithm, close to \textbf{ODDOML}. It
  sends the next block to the
  first worker which is free for computation. As workers
  never have to receive and compute at the same time, the algorithm has no
  extra buffer, so the memory available to store \A, \B, and \C is
  greater. This may change the value of $\mm$ and so the behavior of the
  algorithm.
\end{description}

In the second set we have algorithms which do not use our memory allocation:
\begin{description}
\item[Block Matrix Multiply] (\textbf{BMM}) is Toledo's
  algorithm~\cite{toledo99survey}. It splits each worker memory equally into three parts,
  and allocate one slot for a square block of \A, another for a
  square block of \B, and the last one for a square block of \C, each
  square block having the same size. Then it sends blocks to the workers
  in a demand-driven fashion, when a worker is free for
  computation. First a worker receives a block of \C, then it
  receives corresponding blocks of \A and \B in order to update \C,
  until \C is fully computed. In this version, a worker do not overlap computation
  with the receiving of the next blocks.
\item[Overlapped Block Matrix Multiply] (\textbf{OBMM}) is our attempt to  improve
   the previous algorithm. We try to overlap the
  communications and the computations of the workers. To that purpose, we
  split each worker memory into five parts, so as to receive one block
  of \A and one block of \B while previous ones are used to update \C.
\end{description}

\subsection{Experiments}

We have built several experimental protocols in order to assess the
performance of the various algorithms. In the following experiments we
use nine processors, one master and eight workers. In all experiments
we compare the execution time needed by the algorithms which use our
memory allocation to the execution time of the other algorithms. We
also point out the number of processors used by each algorithm, which
is an important parameter when comparing execution times.

In the first set of experiments, we test the different algorithms on
matrices of different sizes and shapes. The matrices we are
multiplying are of actual size\\
- $8000 \times 8000$ for \A and  $8000 \times 64000$ for $\B$,\\
- $16000 \times 16000$ for \A and  $16000 \times 128000$ for $\B$, and\\
- $8000 \times 64000$ for \A and  $64000 \times 64000$ for $\B$.\\
All the algorithms using our optimized
memory layout consider these matrices as composed of square
blocks of size $q \times q = 80\times 80$. For instance in the first case
we have $\rrr = \ttt = 100$ and $\sss = 800$.


In the second set of experiments we check whether the choice of $q$
was wise. For that purpose, we launch the algorithms on matrices of
size $8000 \times 8000$ and $8000 \times 64000$, changing from one
experiment to another the size of the elementary square blocks. Then
$q$ will be respectively equal to $40$ and $80$. As the global matrix
size is the same in both experiments, we expect both results to be the
same.

In the third set of experiments we investigate the impact of the
worker memory size onto the performance of the algorithms.
In order to have reasonable execution times, we
use matrices of size $16000 \times 16000$ and $16000
\times 64000$, and the memory size will vary from 132MB to 512MB. We
choose these values to reduce side
effects due to the partition of the matrices into blocks of size $\mm q
\times \mm q$.

In the fourth and last set of experiments we check the stability of the
previous results. To that purpose we launch the same
execution five times, in order to determine the maximum gap between two runs.

\subsection{Results and discussion}

\begin{figure}[tbh]
  \begin{minipage}[b]{0.45\linewidth}
    \centerline{\includegraphics[width=\textwidth]{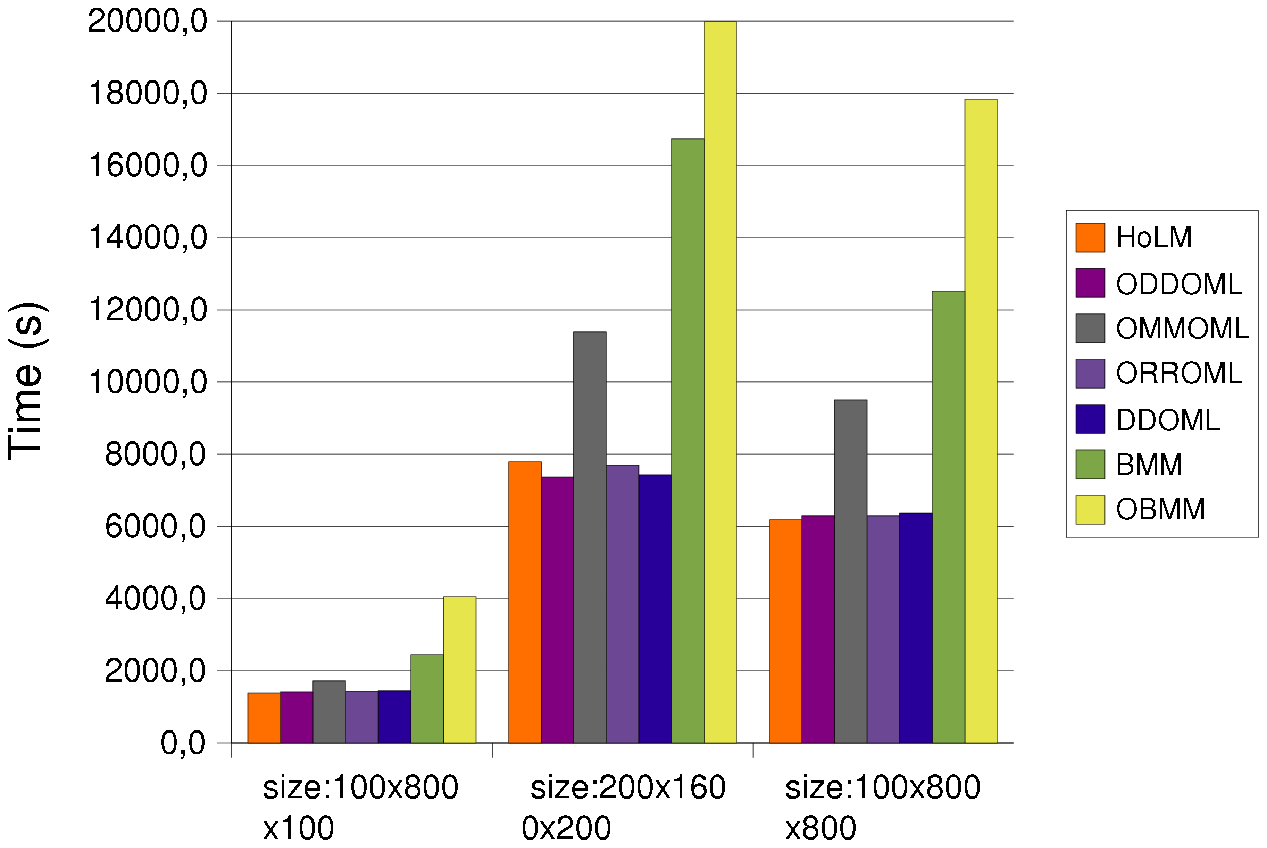}}
    \caption{Performance of the algorithms on different matrices.\label{fig:exp1}}
  \end{minipage}
  \hfill
  \begin{minipage}[b]{0.45\linewidth}
    \centerline{\includegraphics[width=\textwidth]{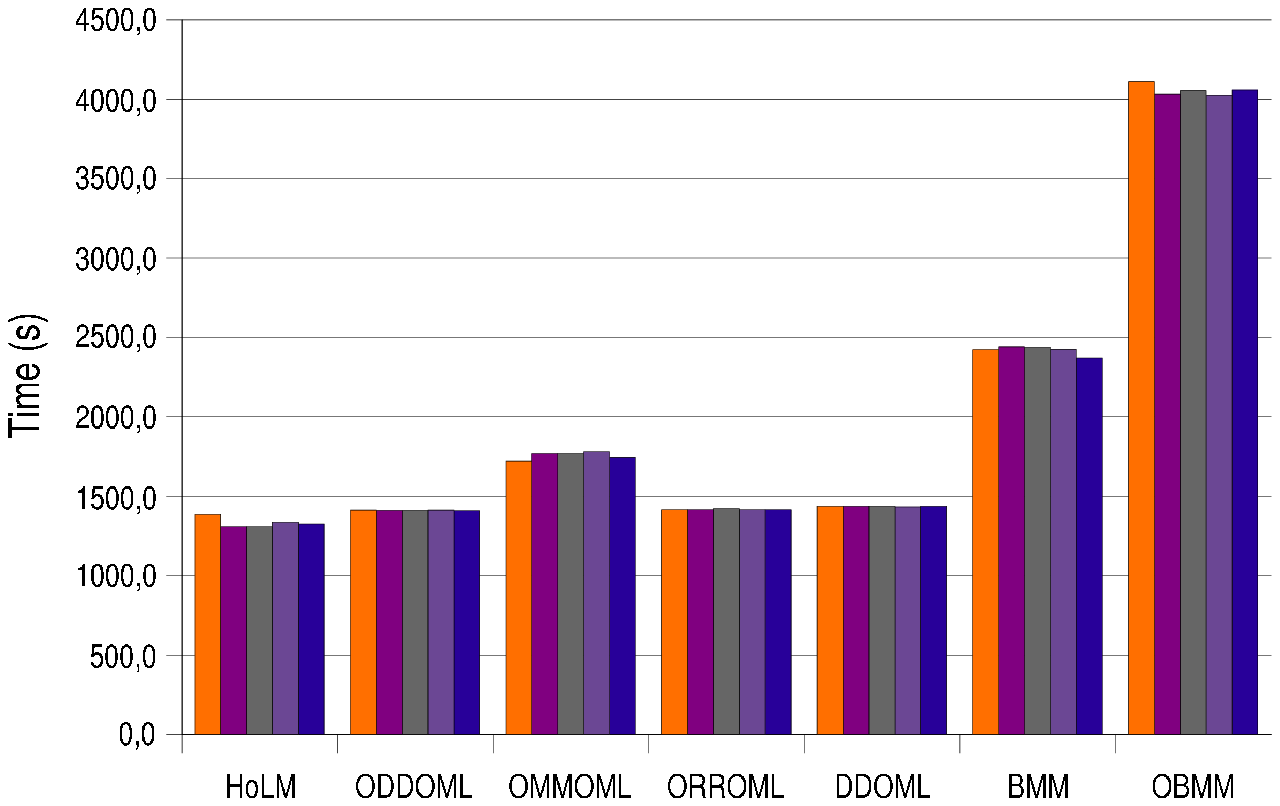}}
    \caption{Variation of algorithm execution times.\label{fig:variation}}
  \end{minipage}
\end{figure}

We see in Figure~\ref{fig:exp1} the results of the first set of
experiments, where algorithms are computing different matrices. The
first remark is that the shape of the three experiments is the same
for all matrix sizes. We also underline the superiority of most of the
algorithms which use our memory allocation against \textbf{BMM}:
\textbf{HoLM}, \textbf{ORROML}, \textbf{ODDOML}, and \textbf{DDOML}
are the best algorithms and have similar performance.
Only \textbf{OMMOML} needs more time to
complete its execution. This delay comes from its resource
selection: it uses only two workers. For instance, \textbf{HoLM} uses four
workers, and is as competitive as the other algorithms which all use the eight
available workers.

\begin{figure}[tbh]
  \begin{minipage}[b]{0.45\linewidth}
    \centerline{\includegraphics[width=\textwidth]{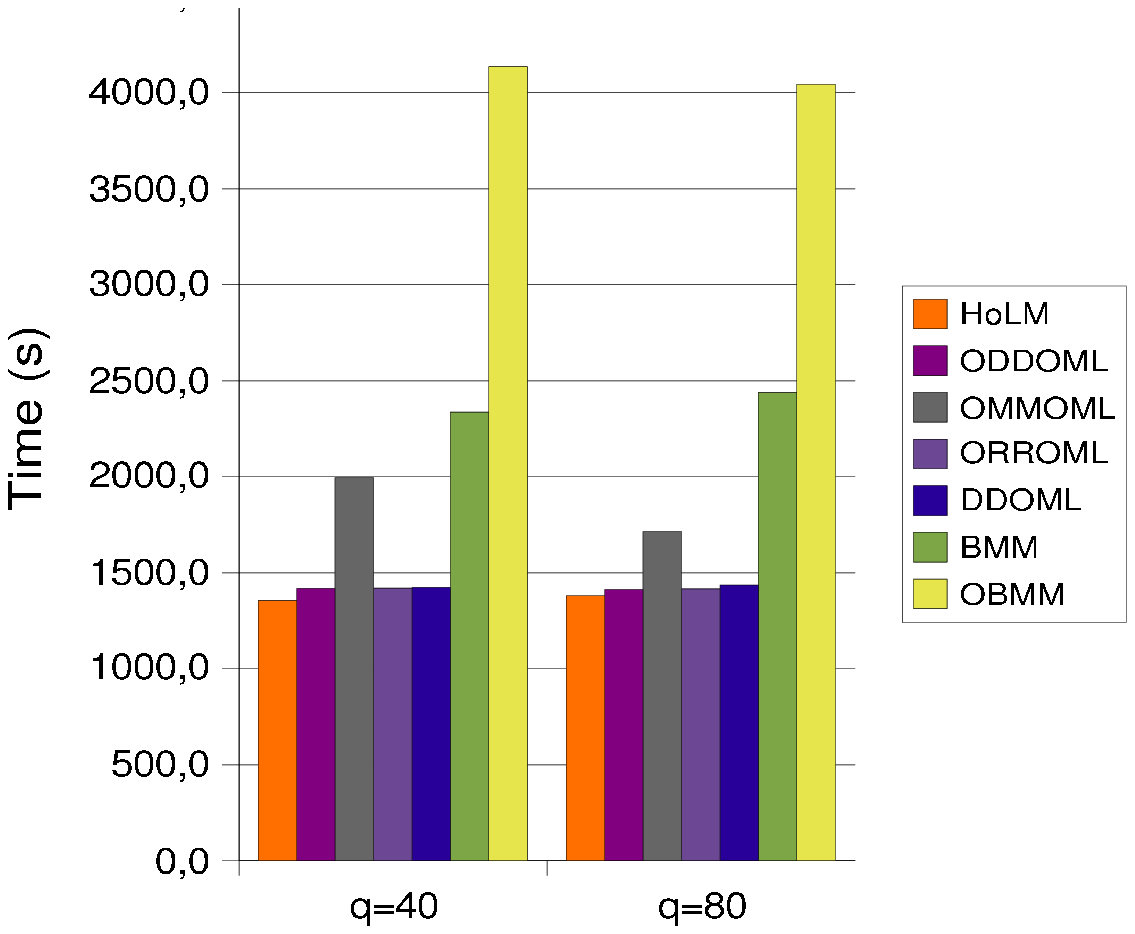}}
    \caption{Impact of block size $q$ on algorithm performance.\label{fig:q}}
  \end{minipage}
  \hfill
  \begin{minipage}[b]{0.45\linewidth}
    \centerline{\includegraphics[width=\textwidth]{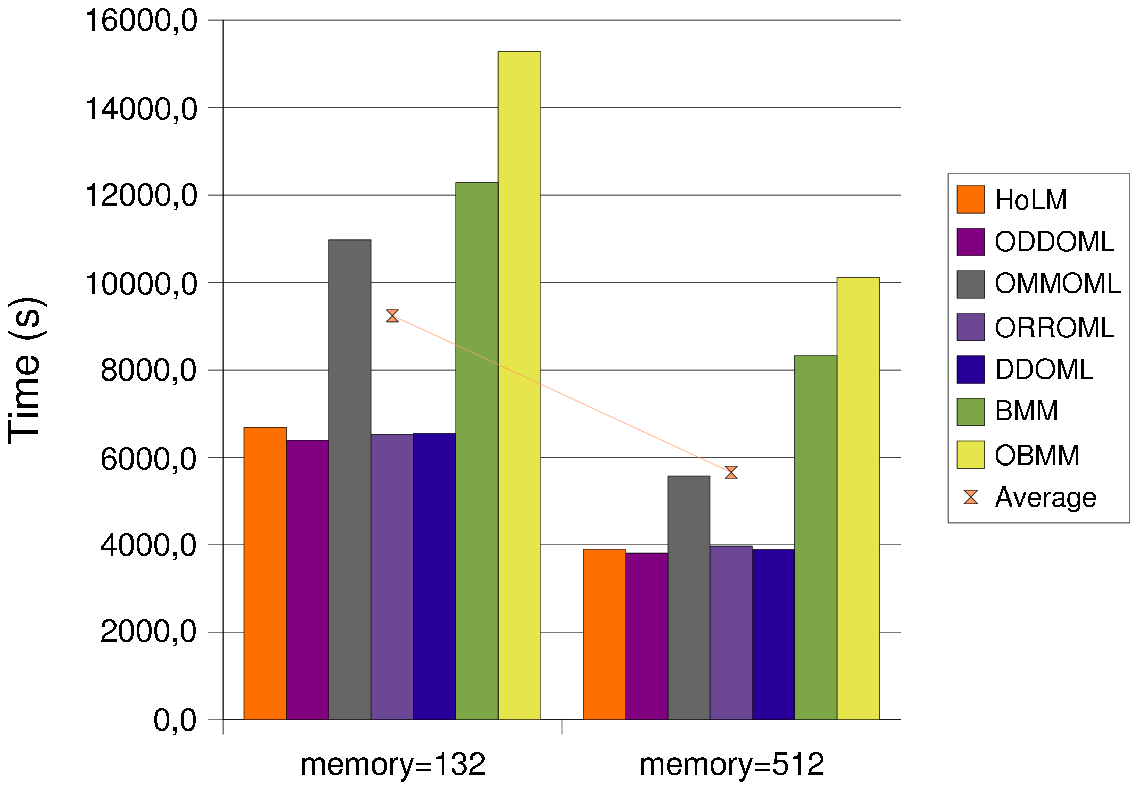}}
    \caption{Impact of memory size on algorithm performance.\label{fig:memory}}
  \end{minipage}
\end{figure}

In Figure~\ref{fig:q}, we see the impact of $q$ on the performance of our algorithms.
\textbf{BMM} and \textbf{OBMM} have same execution times in both experiments as these algorithms
do not split matrices into elementary square blocks of size $q\times q$ but, instead, call the Level 3 BLAS routines directly on the whole $\sqrt{\frac{\mem}{3}}\times\sqrt{\frac{\mem}{3}}$ matrices. In the
two cases we see that the
time of the algorithms are similar.
We point out that this experiment shows that the choice of $q$
has little impact on the algorithms performance.

In Figure~\ref{fig:memory} we have the impact of the worker memory
size on the performance of the algorithms. As expected, the
performance increases with the amount of memory available. It is interesting
to underline that our resource selection always performs in the
best possible way. \textbf{HoLM} will use respectively two and four workers when the memory available increases, compared to
the other algorithms which will use all eight available workers on each test.
\textbf{OMMOML} also makes some resource selection, but it performs worse.

Finally, Figure~\ref{fig:variation} shows the difference that we can
have between two runs. This difference is around $6\%$. Thus if two
algorithms have less than $6\%$ of difference in execution time, they
should be considered as similar.


To conclude, these experiments stress the superiority of our memory
allocation. Furthermore, our homogeneous algorithm is as
competitive as the others but uses fewer resources.

\section{Related work}
\label{sec.related}

In this section, we provide a brief overview of related papers, which
we classify along the following five main lines:

\begin{description}

\item[Load balancing on heterogeneous platforms --]

Load balancing strategies for heterogeneous platforms have been
widely studied. Distributing the computations (together with the
associated data) can be performed either dynamically or statically,
or a mixture of both. Some simple schedulers are available, but they
use naive mapping strategies such as master-worker techniques or
paradigms based upon the idea {\em ``use the past to predict the
future''}, i.e. use the currently observed speed of computation of
each machine to decide for the next distribution of
work~\cite{CiernakZakiLi97a,CiernakZakiLi97b,Berman99}. Dynamic
strategies such as {\em self-guided scheduling}~\cite{Poly88} could
be useful too. There is a challenge in determining a trade-off
between the data distribution parameters and the process spawning
and possible migration policies. Redundant computations might also
be necessary to use a heterogeneous cluster at its best
capabilities. However, dynamic strategies are outside the scope of
this paper (but mentioned here for the sake of completeness).
Because we have a library designer's perspective, we concentrate on
static allocation schemes that are less general and more difficult
to design than dynamic approaches, but which are better suited for
the implementation of fixed algorithms such as linear algebra
kernels from the ScaLAPACK library~\cite{Scalapack97}.

\item[Out-of-core linear algebra routines --] As already mentioned,
the design of parallel algorithms for limited memory processors is
very similar to the design of out-of-core routines for classical parallel
machines. On the theoretical side, Hong and Kung~\cite{HongKung81} investigate the I/O
complexity of several computational kernels in their pioneering paper.
Toledo~\cite{toledo99survey} proposes a nice survey
on the design of out-of-core algorithms for linear algebra, including dense
and sparse computations. We refer to~\cite{toledo99survey} for a complete list of
implementations. The design principles followed by most implementations are
introduced and analyzed by Dongarra et al.~\cite{DongarraHamWal97}.

\item[Linear algebra algorithms on heterogeneous clusters --]

Several authors have dealt with the {\em static} implementation of
matrix-multiplication algorithms on heterogeneous platforms. One
simple approach is given by Kalinov and
Lastovetsky~\cite{KalinovLas99}.  Their idea is to achieve a perfect
load-balance as follows: first they take a fixed layout of
processors arranged as a collection of processor columns; then the
load is evenly balanced {\em within} each processor column
independently; next the load is balanced {\em between} columns; this
is the ``heterogeneous block cyclic distribution''
of~\cite{KalinovLas99}. Another approach is proposed by Crandall and
Quinn~\cite{CrandallQui93}, who propose a recursive partitioning
algorithm, and by Kaddoura, Ranka and Wang~\cite{KaddouraRanWan96},
who refine the latter algorithm and provide several variations. They
report several numerical simulations. As pointed out in the
introduction, theoretical results for  matrix multiplication and LU
decomposition on 2D-grids of heterogeneous processors are reported
in~\cite{ieeeTC2001}, while extensions to general 2D partitioning
are considered in~\cite{ieeeTPDS2001}. See also Lastovetsky and
Reddy~\cite{lasto04} for another partitioning approach.

Recent papers aim at making easier the process of tuning linear algebra kernels on
heterogeneous systems. Self-optimization methodologies are described by Cuenca
et al~\cite{Cuenca05} and by Chen et al~\cite{sans-lfc03}.
Along the same line, Chakravarti et al.~\cite{Lauria2006} describe an implementation of
Cannon's algorithm using self-organizing agents on a peer-to-peer network.

\item[Models for heterogeneous platforms --] In the literature,
one-port models come in two variants. In the unidirectional variant,
a processor cannot be involved in more than one communication at a
given time-step, either a send or a receive. This is the model
that we have used
throughout the paper. In the bidirectional
model, a processor can send and receive in parallel, but at most to
a given neighbor in each direction. In both variants, if $P_u$ sends
a message to $P_v$, both $P_u$ and $P_v$ are blocked throughout the
communication.

The bidirectional one-port model is used by Bhat et
al.~\cite{Bhat99efficient,BhatRagra03} for fixed-size messages. They
advocate its use because ``current hardware and software do not
easily enable multiple messages to be transmitted simultaneously.''
Even if non-blocking, multi-threaded communication libraries allow
for initiating multiple send and receive operations, they claim that
all these operations ``are eventually serialized by the single
hardware port to the network." Experimental evidence of this fact
has recently been reported by Saif and Parashar~\cite{SaifPa2004},
who report that asynchronous MPI sends get serialized as soon as
message sizes exceed a few megabytes. Their results hold for two
popular MPI implementations, MPICH on Linux clusters and IBM MPI on
the SP2.

The one-port model fully accounts for the heterogeneity of the
platform, as each link has a different bandwidth. It generalizes a
simpler model studied by Banikazemi et al.~\cite{Banikazemi98}
Liu~\cite{Liu02} and Khuller and Kim~\cite{khuller04soda}. In this
simpler model, the communication time only depends
on the sender, not on the receiver. In other words, the
communication speed from a processor to all its neighbors is the
same.

Finally, we note that some
papers~\cite{Banikazemi99,barnoy00multicast} depart form the
one-port model as they allow a sending processor to initiate another
communication while a previous one is still on-going on the network.
However, such models insist that there is an overhead time to pay
before being engaged in another operation, so they are not allowing
for fully simultaneous communications.

\item[Master-worker on the computational grid --] Master-worker scheduling
  on the grid can be based on a network-flow
  approach~\cite{ShaoBW00,ShaoPhD} or on an adaptive
  strategy~\cite{HeymannSL00}. Note that the network-flow approach
  of~\cite{ShaoBW00,ShaoPhD} is possible only when using a full
  multiple-port model, where the number of simultaneous communications
  for a given node is not bounded. This approach has also been studied
  in~\cite{hongflow}. Enabling frameworks to facilitate the
  implementation of master-worker tasking are described
  in~\cite{GouxKL00,Weissman00}.

\end{description}

\section{Conclusion}
\label{sec.conclusion}

The main  contributions of this paper are the following:
\begin{enumerate}
\item On the theoretical side, we have derived a new, tighter, bound on
  the minimal volume of communications needed to multiply two matrices.
  From this lower bound, we have defined an efficient memory
  layout, i.e., an algorithm to share the memory
  available on the workers among the three matrices.
\item On the practical side, starting from our memory layout, we have
  designed an algorithm for homogeneous platforms whose performance is
  quite close to the communication volume lower bound. We have extended
  this algorithm to deal with heterogeneous platforms, and discussed how to adapt
  the approach for LU factorization.
\item Through MPI experiments, we have shown
  that our algorithm for homogeneous platforms has far better
  performance than solutions using the memory layout proposed
  in~\cite{toledo99survey}. Furthermore, this static homogeneous
  algorithm has similar performance as dynamic algorithms using the
  same memory layout, but uses fewer processors. It is
  therefore a very good candidate for deploying applications on regular,
  homogeneous platforms.
\end{enumerate}

We are currently conducting experiments to assess the performance of the extension
  of the algorithm for
  heterogeneous clusters.

\bibliographystyle{abbrv}
\bibliography{biblio}

\begin{thebibliography}{10}

\bibitem{Banikazemi98}
M.~Banikazemi, V.~Moorthy, and D.~K. Panda.
\newblock Efficient collective communication on heterogeneous networks of
  workstations.
\newblock In {\em Proceedings of the 27th International Conference on Parallel
  Processing {(ICPP'98)}}. IEEE Computer Society Press, 1998.

\bibitem{Banikazemi99}
M.~Banikazemi, J.~Sampathkumar, S.~Prabhu, D.~Panda, and P.~Sadayappan.
\newblock Communication modeling of heterogeneous networks of workstations for
  performance characterization of collective operations.
\newblock In {\em {HCW'99}, the 8th Heterogeneous Computing Workshop}, pages
  125--133. IEEE Computer Society Press, 1999.

\bibitem{j87}
C.~Banino, O.~Beaumont, L.~Carter, J.~Ferrante, A.~Legrand, and Y.~Robert.
\newblock Scheduling strategies for master-slave tasking on heterogeneous
  processor platforms.
\newblock {\em IEEE Trans. Parallel Distributed Systems}, 15(4):319--330, 2004.

\bibitem{barnoy00multicast}
A.~Bar-Noy, S.~Guha, J.~S. Naor, and B.~Schieber.
\newblock Message multicasting in heterogeneous networks.
\newblock {\em SIAM Journal on Computing}, 30(2):347--358, 2000.

\bibitem{ieeeTC2001}
O.~Beaumont, V.~Boudet, A.~Petitet, F.~Rastello, and Y.~Robert.
\newblock A proposal for a heterogeneous cluster {ScaLAPACK} (dense linear
  solvers).
\newblock {\em IEEE Trans. Computers}, 50(10):1052--1070, 2001.

\bibitem{ieeeTPDS2001}
O.~Beaumont, V.~Boudet, F.~Rastello, and Y.~Robert.
\newblock Matrix multiplication on heterogeneous platforms.
\newblock {\em IEEE Trans. Parallel Distributed Systems}, 12(10):1033--1051,
  2001.

\bibitem{algorithmica}
O.~Beaumont, V.~Boudet, F.~Rastello, and Y.~Robert.
\newblock Partitioning a square into rectangles: {NP}-completeness and
  approximation algorithms.
\newblock {\em Algorithmica}, 34:217--239, 2002.

\bibitem{c122}
O.~Beaumont, L.~Carter, J.~Ferrante, A.~Legrand, L.~Marchal, and Y.~Robert.
\newblock Centralized versus distributed schedulers for multiple bag-of-task
  applications.
\newblock In {\em International Parallel and Distributed Processing Symposium
  {IPDPS'2006}}. IEEE Computer Society Press, 2006.

\bibitem{Berman99}
F.~Berman.
\newblock High-performance schedulers.
\newblock In I.~Foster and C.~Kesselman, editors, {\em The Grid: Blueprint for
  a New Computing Infrastructure}, pages 279--309. Morgan-Kaufmann, 1999.

\bibitem{Bhat99efficient}
P.~Bhat, C.~Raghavendra, and V.~Prasanna.
\newblock Efficient collective communication in distributed heterogeneous
  systems.
\newblock In {\em {ICDCS'99} 19th International Conference on Distributed
  Computing Systems}, pages 15--24. IEEE Computer Society Press, 1999.

\bibitem{BhatRagra03}
P.~Bhat, C.~Raghavendra, and V.~Prasanna.
\newblock Efficient collective communication in distributed heterogeneous
  systems.
\newblock {\em Journal of Parallel and Distributed Computing}, 63:251--263,
  2003.

\bibitem{BlackfordCC96}
L.~Blackford, J.~Choi, A.~Cleary, J.~Demmel, I.~Dhillon, J.~Dongarra,
  S.~Hammarling, G.~Henry, A.~Petitet, K.~Stanley, D.~Walker, and R.~C. Whaley.
\newblock Sca{LAPACK}: {A} portable linear algebra library for
  distributed-memory computers - design issues and performance.
\newblock In {\em Supercomputing '96}. IEEE Computer Society Press, 1996.

\bibitem{Scalapack97}
L.~S. Blackford, J.~Choi, A.~Cleary, E.~D'Azevedo, J.~Demmel, I.~Dhillon,
  J.~Dongarra, S.~Hammarling, G.~Henry, A.~Petitet, K.~Stanley, D.~Walker, and
  R.~C. Whaley.
\newblock {\em {ScaLAPACK} Users' Guide}.
\newblock SIAM, 1997.

\bibitem{Cannon69}
L.~E. Cannon.
\newblock {\em A cellular computer to implement the Kalman filter algorithm}.
\newblock PhD thesis, Montana State University, 1969.

\bibitem{Lauria2006}
A.~Chakravarti, G.~Baumgartner, and M.~Lauria.
\newblock Self-organizing scheduling on the organic grid.
\newblock {\em Int. Journal of High Performance Computing Applications},
  20(1):115--130, 2006.

\bibitem{sans-lfc03}
Z.~Chen, J.~Dongarra, P.~Luszczek, and K.~Roche.
\newblock Self adapting software for numerical linear algebra and lapack for
  clusters.
\newblock {\em Parallel Computing}, 29(11-12):1723--1743, 2003.

\bibitem{CiernakZakiLi97a}
M.~Cierniak, M.~Zaki, and W.~Li.
\newblock Compile-time scheduling algorithms for heterogeneous network of
  workstations.
\newblock {\em The Computer Journal}, 40(6):356--372, 1997.

\bibitem{CiernakZakiLi97b}
M.~Cierniak, M.~Zaki, and W.~Li.
\newblock Customized dynamic load balancing for a network of workstations.
\newblock {\em Journal of Parallel and Distributed Computing}, 43:156--162,
  1997.

\bibitem{CLRbook}
T.~H. Cormen, C.~E. Leiserson, and R.~L. Rivest.
\newblock {\em Introduction to Algorithms}.
\newblock The MIT Press, 1990.

\bibitem{CrandallQui93}
P.~E. Crandall and M.~J. Quinn.
\newblock Block data decomposition for data-parallel programming on a
  heterogeneous workstation network.
\newblock In {\em 2nd International Symposium on High Performance Distributed
  Computing}, pages 42--49. IEEE Computer Society Press, 1993.

\bibitem{Cuenca05}
J.~Cuenca, L.~P. Garcia, D.~Gimenez, and J.~Dongarra.
\newblock Processes distribution of homogeneous parallel linear algebra
  routines on heterogeneous clusters.
\newblock In {\em {HeteroPar'2005}: International Conference on Heterogeneous
  Computing}. IEEE Computer Society Press, 2005.

\bibitem{DongarraHamWal97}
J.~Dongarra, S.~Hammarling, and D.~Walker.
\newblock Key concepts for parallel out-of-core {LU} factorization.
\newblock {\em Parallel Computing}, 23(1-2):49--70, 1997.

\bibitem{GouxKL00}
J.~P. Goux, S.~Kulkarni, J.~Linderoth, and M.~Yoder.
\newblock An enabling framework for master-worker applications on the
  computational grid.
\newblock In {\em {Ninth IEEE International Symposium on High Performance
  Distributed Computing (HPDC'00)}}. IEEE Computer Society Press, 2000.

\bibitem{HeymannSL00}
E.~Heymann, M.~A. Senar, E.~Luque, and M.~Livny.
\newblock Adaptive scheduling for master-worker applications on the
  computational grid.
\newblock In R.~Buyya and M.~Baker, editors, {\em {Grid Computing - GRID
  2000}}, pages 214--227. Springer-Verlag LNCS 1971, 2000.

\bibitem{hongflow}
B.~Hong and V.~Prasanna.
\newblock Bandwidth-aware resource allocation for heterogeneous computing
  systems to maximize throughput.
\newblock In {\em Proceedings of the 32th International Conference on Parallel
  Processing {(ICPP'2003)}}. IEEE Computer Society Press, 2003.

\bibitem{HongKung81}
J.-W. Hong and H.~Kung.
\newblock {I/O} complexity: the red-blue pebble game.
\newblock In {\em {STOC '81: Proceedings of the 13th ACM symposium on Theory of
  Computing}}, pages 326--333. ACM Press, 1981.

\bibitem{toledoJPDC}
D.~Ironya, S.~Toledo, and A.~Tiskin.
\newblock Communication lower bounds for distributed-memory matrix
  multiplication.
\newblock {\em J. Parallel Distributed Computing}, 64(9):1017--1026, 2004.

\bibitem{KaddouraRanWan96}
M.~Kaddoura, S.~Ranka, and A.~Wang.
\newblock Array decomposition for nonuniform computational environments.
\newblock {\em Journal of Parallel and Distributed Computing}, 36:91--105,
  1996.

\bibitem{KalinovLas99}
A.~Kalinov and A.~Lastovetsky.
\newblock Heterogeneous distribution of computations while solving linear
  algebra problems on networks of heterogeneous computers.
\newblock In P.~Sloot, M.~Bubak, A.~Hoekstra, and B.~Hertzberger, editors, {\em
  HPCN Europe 1999}, LNCS 1593, pages 191--200. Springer Verlag, 1999.

\bibitem{khuller04soda}
S.~Khuller and Y.~Kim.
\newblock On broadcasting in heterogenous networks.
\newblock In {\em Proceedings of the fifteenth annual ACM-SIAM symposium on
  Discrete algorithms}, pages 1011--1020. Society for Industrial and Applied
  Mathematics, 2004.

\bibitem{lasto04}
A.~Lastovetsky and R.~Reddy.
\newblock Data partitioning with a realistic performance model of networks of
  heterogeneous computers.
\newblock In {\em International Parallel and Distributed Processing Symposium
  {IPDPS'2004}}. IEEE Computer Society Press, 2004.

\bibitem{Liu02}
P.~Liu.
\newblock Broadcast scheduling optimization for heterogeneous cluster systems.
\newblock {\em Journal of Algorithms}, 42(1):135--152, 2002.

\bibitem{MaheswaranAS99}
M.~Maheswaran, S.~Ali, H.~Siegel, D.~Hensgen, and R.~Freund.
\newblock Dynamic matching and scheduling of a class of independent tasks onto
  heterogeneous computing systems.
\newblock In {\em Eight Heterogeneous Computing Workshop}, pages 30--44. IEEE
  Computer Society Press, 1999.

\bibitem{Poly88}
C.~D. Polychronopoulos.
\newblock Compiler optimization for enhancing parallelism and their impact on
  architecture design.
\newblock {\em IEEE Transactions on Computers}, 37(8):991--1004, Aug. 1988.

\bibitem{SaifPa2004}
T.~Saif and M.~Parashar.
\newblock Understanding the behavior and performance of non-blocking
  communications in {MPI}.
\newblock In {\em Proceedings of Euro-Par 2004: Parallel Processing}, LNCS
  3149, pages 173--182. Springer, 2004.

\bibitem{ShaoPhD}
G.~Shao.
\newblock {\em Adaptive scheduling of master/worker applications on distributed
  computational resources}.
\newblock PhD thesis, {Dept. of Computer Science, University Of California at
  San Diego}, 2001.

\bibitem{ShaoBW00}
G.~Shao, F.~Berman, and R.~Wolski.
\newblock Master/slave computing on the grid.
\newblock In {\em Heterogeneous Computing Workshop {HCW'00}}. IEEE Computer
  Society Press, 2000.

\bibitem{toledo99survey}
S.~Toledo.
\newblock A survey of out-of-core algorithms in numerical linear algebra.
\newblock In {\em External Memory Algorithms and Visualization}, pages
  161--180. American Mathematical Society Press, 1999.

\bibitem{Weissman00}
J.~B. Weissman.
\newblock Scheduling multi-component applications in heterogeneous wide-area
  networks.
\newblock In {\em Heterogeneous Computing Workshop {HCW'00}}. IEEE Computer
  Society Press, 2000.

\bibitem{atlas_sc98}
R.~C. Whaley and J.~Dongarra.
\newblock Automatically tuned linear algebra software.
\newblock In {\em Proceedings of the ACM/IEEE Symposium on Supercomputing
  (SC'98)}. IEEE Computer Society Press, 1998.

\end{thebibliography}

\end{document}